\documentclass[11pt,a4paper]{article}

\usepackage[symbol]{footmisc}

\usepackage{jheppub}

\usepackage{slashed}
\usepackage{diagbox}
\usepackage{subfigure}
\usepackage{amsmath}
\usepackage{dutchcal}
\usepackage{mathrsfs}
\usepackage{amssymb}
\usepackage{bbm}

\usepackage{amsthm}
\usepackage{amsfonts}
\usepackage{dsfont}
\usepackage{amssymb, bm}
\usepackage{array}
\usepackage{booktabs}
\usepackage{overpic}

\def\II{\hbox{{1}\kern-.25em\hbox{l}}}

\newcommand{\beq}{\begin{equation}}
\newcommand{\eeq}{\end{equation}}
\newcommand{\bal}{\begin{align}}
\newcommand{\eal}{\end{align}}
\newcommand{\x}{{\bm x}}
\newcommand{\y}{{\bm y}}

\newcommand{\nn}{\nonumber}

\newcommand{\scr}{\mathscr}
\newcommand{\bb}{\mathbb}

\newcommand \widebar [1] {\overline{#1}}

\newcommand{\parallelsum}{\mathbin{\!/\mkern-5mu/\!}}

\def \e {\mbox{e}}
\def \qqquad {\qquad\quad}

\makeatletter
\newcommand{\newparallel}{\mathrel{\mathpalette\new@parallel\relax}}
\newcommand{\new@parallel}[2]{%
  \begingroup
  \sbox\z@{$#1T$}
  \resizebox{!}{\ht\z@}{\raisebox{\depth}{$\m@th#1/\mkern-5mu/$}}%
  \endgroup
}
\makeatother

\allowdisplaybreaks

\newcounter{MBQ}

\def\correspondingauthor{\footnote{Corresponding author.}}

\title{Connecting Euclidean to light-cone correlations:
From flavor nonsinglet in forward kinematics to flavor singlet in non-forward kinematics}

\author[a]{Fei Yao,}

\author[b]{Yao Ji\correspondingauthor{}}

\author[c,a]{and Jian-Hui Zhang\correspondingauthor{}}

\subheader{
\begin{flushright}
{\small
TUM-HEP-1446/22\\
}
\end{flushright}
}

\affiliation[a]{
   Center of Advanced Quantum Studies, Department of Physics, Beijing Normal University, Beijing 100875, China}

\affiliation[b]{
   Physik Department T31, James-Franck-Stra\ss e 1, 
   Technische Universit{\"a}t M{\"u}nchen,\\
   D-85748 Garching, Germany}
 
\affiliation[c]{School of Science and Engineering, The Chinese University of Hong Kong, Shenzhen 518172, China}

\emailAdd{feiyao@mail.bnu.edu.cn}

\emailAdd{yao.ji@tum.de} 

\emailAdd{zhangjianhui@cuhk.edu.cn}

\abstract{
We present a unified framework for the perturbative factorization connecting Euclidean correlations to light-cone correlations. Starting from nonlocal quark and gluon bilinear correlators, we derive the relevant hard-matching kernel up to the next-to-leading-order, both for the flavor singlet and non-singlet combinations, in non-forward and forward kinematics, and in coordinate and momentum space.
The results for the generalized distribution functions (GPDs), parton distribution functions (PDFs), and distribution amplitudes (DAs) are obtained by choosing appropriate kinematics. The renormalization and matching are done in a state-of-the-art scheme. We also clarify some issues raised on the perturbative matching of GPDs in the literature. Our results provide a complete manual for extracting all leading-twist GPDs, PDFs 
as well as DAs from lattice simulations of Euclidean correlations in a state-of-the-art strategy, either in coordinate or in momentum space factorization approach.
}


\begin{document}

\maketitle


\section{Introduction}
\label{SEC:Introduction}

The partonic structure of hadrons plays a crucial role in mapping out their 3D image and in describing the experimental data collected at high energy colliders such as the Large Hadron Collider (LHC) and the Electron-Ion Collider (EIC). The simplest quantities that characterize such structure include the leading-twist collinear parton distribution functions (PDFs), generalized parton distributions (GPDs) and distribution amplitudes (DAs). They are all defined by collinear parton operators and are nonperturbative in nature. A lot of efforts have been devoted to determining them from fitting various experimental data for both singlet and non-singlet sectors~(see, e.g., \cite{Hou:2019efy,PDF4LHCWorkingGroup:2022cjn,Kumericki:2016ehc,Lin:2020rut} and references therein). Among them, the experimental extraction of the leading-twist GPDs is much more difficult due to their complicated kinematic dependence. 
Some progress has been made on reducing the uncertainties in perturbative QCD~\cite{Braun:2020yib,Gao:2021iqq,Braun:2021grd, Braun:2022bpn} and from power corrections~\cite{Braun:2012hq, Braun:2014sta, Braun:2014paa,Braun:2020zjm,Braun:2022qly}.

{In the past few years, 
we have also witnessed rapid developments} on extracting these quantities from lattice 
QCD~(see \cite{Cichy:2018mum,Ji:2020ect,Constantinou:2020hdm,Constantinou:2022yye} for a recent review) based on various proposals~\cite{Liu:1993cv,Braun:2007wv,Ji:2013dva,Ji:2014gla,Ma:2017pxb,Lin:2017snn,Radyushkin:2017cyf,Detmold:2005gg,Chambers:2017dov}. Among them, one of the widely used options is to start from the quasi-light-front (quasi-LF) correlators~\cite{Ji:2013dva,Ji:2014gla,Ji:2020ect}, which are equal-time quark and gluon correlators defined on a Euclidean space interval. 
It can be connected to the light-front (LF) correlators defining the collinear parton distributions, either through a short-distance factorization~\cite{Radyushkin:2017cyf} in coordinate space or through a large-momentum factorization~\cite{Ji:2013dva,Xiong:2013bka,Ji:2020ect} in momentum space. Such a factorization requires a hard coefficient function or matching kernel as the input, which is, in general, different {for different hadronic functions.}

There have been many studies on the perturbative matching relevant to the computation of collinear parton 
{distributions}~(for a summary, see Refs.~\cite{Ji:2020ect,Constantinou:2022yye}), most of which are focused on the flavor-nonsinglet combination due to its simplicity. The state-of-the-art is the next-to-next-to-leading order (NNLO) calculation for the isovector unpolarized quark PDFs~\cite{Chen:2020ody,Li:2020xml}. However, the majority of the results cannot be directly implemented in lattice calculations. This is because, on the one hand, dimensional regularization (DR) and $\overline{\rm MS}$ scheme are often used in such calculations, while these schemes cannot be realized in lattice calculations. Therefore, a conversion between the lattice renormalization scheme to the $\overline{\rm MS}$ scheme needs to be performed before the matching can be applied. On the other hand, certain schemes are regularization-independent~\cite{Chen:2016fxx,Izubuchi:2018srq,Alexandrou:2017huk,Radyushkin:2018cvn}, and thus avoid the scheme conversion mentioned above. But in such schemes the renormalization factors, when applied in the large-momentum factorization approach, introduce undesired infrared (IR) contributions at large distances, and thus invalidate the factorization formula. In Ref.~\cite{Ji:2020brr}, a hybrid renormalization scheme has been proposed to circumvent this issue, and used in subsequent lattice calculations of PDFs and DAs~\cite{Hua:2020gnw,Gao:2021dbh,LatticeParton:2022zqc,Gao:2022vyh,Gao:2022iex,LatticeParton:2022xsd,Gao:2022uhg}. 

In contrast to the flavor-nonsinglet combination, the flavor-singlet quark and gluon combinations have been much less studied~\cite{Wang:2017qyg,Wang:2017eel,Wang:2019tgg,Balitsky:2019krf,Balitsky:2021cwr,Balitsky:2021qsr} in the literature. In Refs.~\cite{Wang:2019tgg,Balitsky:2019krf}, the results for the unpolarized flavor-singlet quark and gluon PDFs have been presented, where a discrepancy was found in the unphysical region for the matching kernel of the unpolarized gluon PDF. In Ref.~\cite{Ma:2022ggj}, the one-loop matching for quark GPDs has been recalculated using a subtraction approach, in which a discrepancy from previous calculations~\cite{Ji:2015qla,Xiong:2015nua} was found in the flavor-nonsinglet combination. In the same paper, certain mixing contributions between the flavor-singlet quark GPDs and the gluon GPDs have also been considered. However, a complete calculation of both coordinate and momentum space factorizations in the state-of-the-art renormalization scheme is still missing.

Given that the PDFs, DAs and GPDs are all defined by the same collinear parton operators, it is highly desirable to develop a unified computational framework for all of them, so that the perturbative matching for each can be obtained by a choice of appropriate kinematics. It is the purpose of the present paper to develop such a framework. 
We start from nonlocal quark and gluon bilinear operators, and first present a general coordinate space factorization formula. The relevant coordinate space matching kernels are then calculated for general non-forward kinematics, both for the flavor singlet and non-singlet combinations, 
where the renormalization and matching are done in a state-of-the-art scheme. We also clarify some of the observed discrepancies between results by different groups. Our results provide a complete manual for a state-of-the-art extraction of all leading-twist GPDs, PDFs 
as well as DAs from lattice simulations, either in coordinate or in momentum space factorization approach.  
Our framework will also greatly facilitate higher-order perturbative calculations involving collinear parton correlators of spacelike separation, which are important for extracting collinear parton structure functions from lattice calculations to a high accuracy.

The rest of the paper is organized as follows: In Sec.~\ref{SEC:facteucllf}, we give the operator definitions used throughout the paper, and present a factorization between quasi-LF and LF correlators that is valid in general kinematics, both in coordinate space and in momentum space. In Sec.~\ref{SEC:matchnonforward}, we show the calculation of the matching kernel for flavor-singlet quark and gluon combinations in non-forward kinematics, and give the results in the $\overline{\rm MS}$, ratio and hybrid schemes. We then discuss in Sec.~\ref{SEC:pdfdalimits} the special kinematic limits in which the PDFs and DAs are recovered, and present the corresponding matching kernels in all schemes. Sec.~\ref{SEC:conclusion} is our conclusion. Some technical details of the calculation are summarized in the Appendix.

\section{Factorization between Euclidean and light-front correlations}
\label{SEC:facteucllf}

In this section, we give the definition of the relevant quasi-LF correlators, and present the factorization formula connecting them to the LF correlators defining collinear parton distributions.

\subsection{Operator {bases}}
\label{SEC:operator}

To begin with, let us consider the following spatial non-local quark correlator
\beq
O_{q}(z_1, z_2)=\bar \psi(z_1)\Gamma [z_1, z_2] \psi(z_2),
\eeq
where $\Gamma$ is the Dirac structure taken as $\gamma^t, \gamma^z\gamma_5, \gamma^t\gamma^\perp\gamma_5$ in the present paper. They are used to define the unpolarized, longitudinally polarized (helicity) and transversely polarized (transversity) quark quasi-LF correlations, respectively. In the discussion below, we will denote them by subscripts $u, h, t$, respectively. $z_1, z_2$ are 4-vectors along the spatial $z$ direction, and $[z_1, z_2]$ represents the straight Wilson line in the fundamental representation ensuring gauge invariance of the operator, 
\beq
[z_1, z_2]= {\text P} \exp\bigg[ig\int_{0}^{1}dt\, z_{12}\cdot A(z_2+ t z_{12})\bigg]
\eeq
with $z_{12}^\mu=z_1^\mu-z_2^\mu={\bf{z}}_1 v^\mu-{\bf{z}}_2 v^\mu={\bf{z}}_{12} v^\mu$ ($v^2=-1$) and ${\text P}$ indicating the path-ordering, 
\begin{eqnarray}\label{def:gamma5}
\gamma_5=i\gamma^0\gamma^1\gamma^2\gamma^3=\frac{-i}{4!}\epsilon^{\mu\nu\rho\sigma}\gamma_\mu\gamma_\nu\gamma_\rho\gamma_\sigma,
\end{eqnarray}
where $\epsilon^{\mu\nu\rho\sigma}$ is the Levi-Civita tensor with $\epsilon^{0123}=-\epsilon_{0123}=1$. 
Flavor indices of the quark fields are omitted for brevity. 

{The flavor non-singlet and singlet quark combinations 
are related to the following combinations of quark correlators (summation over quark flavors will be omitted hereafter for simplicity)
}
\begin{align}\label{def:quark-op}
	{O}^{ns}_{q}(z_1,z_2) &=\frac{1}{2} \left[ O_{q}(z_1,z_2) \pm O_{q}(z_2,z_1)\right], 
\  {O}_{q}^s(z_1,z_2) =\frac{1}{2} \left[ O_{q}(z_1,z_2) \mp O_{q}(z_2,z_1) \right],
\end{align}
where the upper sign corresponds to the unpolarized {(vector)} case while the lower sign corresponds to the helicity {(axial-vector)} case. Note that the operator defining the quark transversity is chiral-odd and thus does not mix with gluons, and the combination of the corresponding quark correlators always takes ``+" sign.

Analogously, we can introduce the following spatial nonlocal operators for gluons in $d$ spacetime dimensions,
\begin{align}\label{eq:gluonope}
	{O}_{g,u}(z_1,z_2) &= g^{\mu\nu}_{{\perp}} {\bf F}_{\mu\nu}\,, \notag\\
	{O}_{g,h}(z_1,z_2)  &=i \epsilon_\perp^{\mu\nu} {\bf F}_{\mu\nu}\, ,
	\notag\\
	{O}_{g,t}\,(z_1,z_2) &=\hat{S}\,{\bf F}_{\mu\nu}=\frac{1}{2}\left[{\bf F}_{\mu\nu}+{\bf F}_{\nu\mu}\right]-\frac{1}{d-2}\, g_\perp^{\mu\nu} \,{{\bf F}_\alpha} ^\alpha  ,
\end{align}
where ${\bf F}_{\mu\nu} = z_{12}^\rho {\rm F}_{\rho\mu}(z_1) [z_1, z_2]{\rm F}_{\nu\sigma}(z_2) z_{12}^\sigma \equiv\text F_{z_{12}\mu} (z_1) [z_1,z_2] \,\text F_{\nu z_{12}} (z_2)$ with ${\rm F}_{\mu\nu}$ being the gluon field strength tensor. The Wilson line $[z_1, z_2]$ in this case resides in the adjoint representation. {$\hat{S}$ indicates a symmetrization over the indices and a trace subtraction in the operator that follows}. $\mu,\nu,\alpha$ represent the $d-2$ dimensional transverse indices in DR. In lattice simulations, $d$ should be replaced by 4. In $O_{g,u}$ the summation over $\mu,\nu$ can also be taken over all Lorentz components (i.e., including the longitudinal and time directions), which generates identical leading-twist LF distributions. We also consider this operator choice and list the results in the Appendix. {Note that Bose symmetry indicates that the gluon operators must satisfy certain relations under $z_1\leftrightarrow z_2$. For example, $O_{g,h}(z_1,z_2)$ shall be anti-symmetric, whereas $O_{g,\{u,t\}}(z_1,z_2)$ shall be symmetric under the exchange $z_1\leftrightarrow z_2$}.

\subsection{Quasi-light-front correlations}
By sandwiching the nonlocal parton operators~\eqref{def:quark-op} and~\eqref{eq:gluonope} in different external hadronic states, we can define the quasi-LF correlations (qLFCs) relevant for the calculation of GPDs, PDFs and DAs. For example, the quasi-GPDs are defined through the following non-forward matrix element
\begin{align}\label{eq:qLFcorr}
{\cal {H}}_{q,g} (z_i,P_i,\mu)&=\langle P_1 S_1|O_{q,g}(z_1, z_2)|P_2 S_2\rangle, 
\end{align}
where the hadron state is denoted by its momentum and spin as $|P_iS_i\rangle$ with $i=1,2$. The forward quasi-PDF limits are obtained when the initial and final states are identical. Quasi-DAs are hadron-to-vacuum matrix elements of the same operators, and thus can also be obtained from the above matrix element by taking appropriate kinematic limits. 

The general factorization formula of quasi-LF correlators can be summarized into a matrix form as
\begin{align}\label{eq:fac_coord}
	\begin{pmatrix}
		O_q \\[1.5mm]
		O_g
	\end{pmatrix}
	= 
	\begin{pmatrix}
		C_{qq} & C_{qg} \\[1mm]
		C_{gq} & C_{gg}
	\end{pmatrix}
        \otimes
	\begin{pmatrix}
		{O^{l.t.}_q} \\[1.5mm]
		{O^{l.t.}_g} 
	\end{pmatrix}+h.t.\, ,
\end{align}
where we have suppressed the factorization scale dependence for notational simplicity. $l.t.$ stands for the leading-twist projection of the non-local correlator which acts as the generating function of leading-twist local operators~\cite{Balitsky:1987bk,Anikin:1978tj, Muller:1994ses, Balitsky:1990ck, Geyer:1999uq, Braun:2018brg}. The existence of the local operator product expansion naturally implies the factorization formula above. $\otimes$ stands for the convolution, and $h.t.$ denotes higher-twist terms. The $2\times 2$ mixing matrix $C$ is the perturbative matching function in coordinate space, and for the non-singlet combination only $C_{qq}$ is non-vanishing. 
In general, the mixing matrix depends on two Feynman parameters $\alpha, \beta$ and the invariant distance squared $z_{12}^2$, and the convolution takes the form of a double integral. For example, writing down the first row of Eq.~(\ref{eq:fac_coord}) explicitly, we have
{\begin{align}
	O_q(z_1,z_2)= \int_0^1 d\alpha \int_{0}^{\bar{\alpha}} &d\beta\, \Big[C_{qq}(\alpha,\beta,\mu^2z_{12}^2)O_q^{l.t.}(z_{12}^\alpha, z_{21}^\beta)+C_{qg}(\alpha,\beta,\mu^2 z_{12}^2)O_g^{l.t.}(z_{12}^\alpha, z_{21}^\beta)\notag\\
    &+\widetilde{C}_{qq}(\alpha,\beta,\mu^2z_{12}^2)O_q^{l.t.}(z_{21}^\alpha, z_{12}^\beta)+\widetilde{C}_{qg}(\alpha,\beta,\mu^2z_{12}^2)O_g^{l.t.}(z_{21}^\alpha, z_{12}^\beta)\Big],
\end{align}}
where $z^\alpha_{12}=\bar{\alpha}z_1+\alpha z_2$ with $\bar{\alpha}=1-\alpha$. Note that while $\widetilde{C}_{qq}(\alpha,\beta,\mu^2z_{12}^2)$ and $\widetilde{C}_{qg}(\alpha,\beta,\mu^2z_{12}^2)$ vanish at one-loop, they will contribute starting from two-loop level.
The factorization in Eq.~(\ref{eq:fac_coord}) is valid to all orders in perturbation theory. In this work, we lay the foundation of the computational framework and focus on the mixing matrix up to the next-to-leading-order (NLO), and leave higher-order calculations to future publications.

The leading-twist operators $O^{l.t.}$ on the r.h.s. of Eq.~(\ref{eq:fac_coord}) define the leading-twist parton distributions when sandwiched between suitable external hadronic states. {For example, {from the LF quark operator $O^{l.t.}_{\gamma^+}(z_1^-, z_2^-)\equiv \bar \psi(z_1n_-)\gamma^+ [z_1,z_2] \psi(z_2n_-)$  with $n_-^\mu$ being a lightlike vector, we define the leading-twist unpolarized quark GPDs $H$ and $E$~\cite{Muller:1994ses,Ji:1996ek,Radyushkin:1997ki}}
\begin{align}
&\langle P_1 S_1|O^{l.t.}_{\gamma^+}(z_1^-, z_2^-)|P_2 S_2\rangle\nn\\
&=\int_{-1}^1 dx\, e^{i(x+\xi) P^+z_1^- -i(x-\xi)P^+ z_2^-}
\bar u(P_1 S_1)\big[H(x,\xi,t)\gamma^+ +E(x,\xi,t) \frac{i\sigma^{+\mu}\Delta_\mu}{2M} \big]u(P_2 S_2)
\end{align}
 with the kinematic variables
\beq
P^\mu=\frac{P_1^\mu+P_2^\mu}{2}, \ \ \ \ {\Delta^\mu=P_1^\mu-P_2^\mu}, \ \ \ \ t=\Delta^2, \ \ \ \ \xi=\frac{P_1^+-P_2^+}{P_1^+ + P_2^+}.
\eeq
The skewness parameter $\xi$ differs in the qLFCs and LF correlations only by power-suppressed contributions~\cite{Liu:2019urm}. We will not distinguish them in the following.  
The definitions of all leading-twist parton distributions (GPDs, PDFs, DAs) can be found in Refs.~\cite{Diehl:2003ny, Belitsky:2005qn}.

We would like to stress that the matching kernels for the GPDs reduce to that of the PDFs and DAs in certain kinematic limits. Thus, in the following, we will begin with the generic non-forward matrix elements relevant to the GPDs. The PDF and DA limits will then be discussed as special cases.

\subsection{Pseudo-distributions}
One way to extract collinear parton distributions, or more precisely their moments, from the qLFCs is to use the short-distance factorization~\cite{Radyushkin:2018cvn}. Keeping $z_{12}^2$ fixed and Fourier transforming the quasi-LF distance (or Ioffe-time) $\zeta_i=z_i P_z(i=1,2)$ to the momentum fraction, one obtains the so-called pseudo-distributions. We begin with the pseudo-GPDs (pGPDs)\footnote{Throughout this work, we use the calligraphic letter to denote both the LF/Ioffe-time distribution and pseudo-distributions with the former (latter) in coordinate (momentum fraction) space, whereas the quasi-distributions are represented by {{blackboard bold letters}}.}, which are related to the non-forward matrix element defined in the previous subsection via the following Fourier transform
\begin{align}\label{eq:pseudoFT}
	\mathcal P\left(\tau,\,\xi,\,\mu^2 z_{12}^2 \right)=N\int\frac{d \zeta_1}{2\pi}\int\frac{d \zeta_2}{2\pi}\,e^{i(\xi+\tau)\zeta_1+i(\xi-\tau)\zeta_2 }\,{\cal {H}} (\zeta_i,\mu^2 z_{12}^2),
\end{align}
where $N$ is a normalization factor. For the unpolarized, longitudinally and transversely polarized quark pGPDs $N=1/(2P_t)$, $1/(2P_z)$ and $1/(2P_tS_\perp)$ respectively, while for gluon pGPDs $N=1$ regardless of the polarization.
The pGPDs can be factorized into the GPDs as follows
  \begin{align}\label{eq:pseudoFactor}
 	&\mathcal P\left(\tau,\xi,{\mu^2 z_{12}^2}\right)= \int_{-1}^{1}dy\, \mathcal{C}\left(\tau_1,\tau_2 , y_1,y_2;{\mu^2 z_{12}^2}\right) H(y,\xi,\mu), 	
 \end{align}
 with the $2\times 2$ matching matrix 
\begin{align}
\mathcal{C}\left(\tau_1,\tau_2 , y_1,y_2;\mu^2 z_{12}^2\right)
= 
\begin{pmatrix}
	\mathcal{C}_{qq} & \mathcal{C}_{qg} \\[1mm]
	\mathcal{C}_{gq} & \mathcal{C}_{gg}
\end{pmatrix} \,,
\end{align}
where we follow the notation of Ref.~\cite{Belitsky:2005qn}, $\tau_1$, $\tau_2$, $y_1$ and $y_2$ are momentum fractions that can be expressed in a more symmetric way as
\begin{align}
	\tau_1=\xi+\tau\,, \qquad \tau_2=\xi-\tau\, , \qquad y_1=\xi+y\, , \qquad y_2=\xi-y\, .\notag
\end{align} 
The matching kernels in coordinate- and pseudo-space are related by the Fourier transform~\eqref{eq:pseudoFT}, which leads to the following formula that allows us to get the latter from the former directly,
\begin{align}~\label{eq:qLFC2ps}
	\mathcal{C}(\tau_1,\tau_2 , y_1,y_2;\mu^2 z_{12}^2)
	= \int_{0}^{1} d\alpha \int_{0}^{1} d\beta\, C(\alpha,\beta,{\mu^2 z_{12}} ) \,\delta(\tau_1-\bar{\alpha}y_1-\bar{\alpha}\beta y_2),
\end{align}
where a delta function enforcing the momentum conservation $\tau_1 + \tau_2 = y_1 + y_2$ is omitted.
{Two formulas prove to be helpful for computing the integral above. we include them in the Appendix together with some technical details. It is worth pointing out that the symmetry relations for the diagonal and off-diagonal elements ${\cal{C}}_{aa}(\tau, y) = {\cal{C}}_{aa}(-\tau, -y)$, ${\cal{C}}_{ab}(\tau, y) = -{\cal{C}}_{ab}(-\tau, -y)$ are recovered automatically from the integral in~\eqref{eq:qLFC2ps} following the operator definitions~\eqref{def:quark-op},~\eqref{eq:gluonope} as expected.
}

\subsection{Quasi-distributions}
In the large-momentum factorization or large-momentum effective theory (LaMET)~\cite{Ji:2013dva,Ji:2014gla,Ji:2020ect} approach, one defines the quasi-distribution as a Fourier transform of the qLFCs with respect to the spatial interval. 
For the purpose of deriving the matching kernel for quasi-distributions, we can either Fourier transform the qLFCs directly to momentum space, 
or first to pseudo-space and then do a double Fourier transform to momentum space as adopted in Refs.~\cite{Braun:2021aon,Braun:2021gvv}. Both strategies yield the same result, but the second one turns out to be easier, so we use the second strategy for our calculation. 
The quasi-GPDs (qGPDs) are then related to the pGPDs as 
 \beq\label{eq:quasiGPDdef}
	\bb H\left(x,\xi,{{\frac{\mu}{P_z}}}\right) =\int^1_{-1}d\tau_1 \int^1_{-1}d\tau_2 \int \frac{d \zeta_1}{2\pi} \int \frac{d \zeta_2}{2\pi} \, e^{i\left[(x_1-\tau_1)\zeta_1+(x_2-\tau_2)\zeta_2 \right]}\, 	\mathcal {P} \left(\tau_1,\tau_2,\frac{\mu^2 \zeta_{12}^2}{P_z^2}\right)  ,
\eeq
where $x_1=\xi+x$, $x_2=\xi-x$ and $\zeta_{12}=\zeta_1-\zeta_2$, and the factorization formula for the qGPDs reads
\begin{align}\label{eq:qGPDfact}
 \bb {H}\left(x,\xi,{{\frac{\mu}{P_z}}}\right)&= \int_{-1}^{1}dy\, 
 \bb {C}\left(x_1,x_2, y_1,y_2;{{\frac{\mu}{P_z}}}\right) H(y,\xi,\mu),	
 \end{align}
 with the matching coefficient in momentum space
\begin{align}
\bb{C}\left(x_1,x_2 , y_1,y_2;{{\frac{\mu}{P_z}}}\right)
= 
\begin{pmatrix}
	\bb{C}_{qq} & \bb{C}_{qg} \\[1mm]
	\bb{C}_{gq} & \bb{C}_{gg}
\end{pmatrix} \,.
\end{align} 
Since the $z_{12}$-dependence in the pGPDs only appears through logarithmic functions, the $z_i$-integrals in~\eqref{eq:quasiGPDdef} can be computed in a procedural manner~\cite{Braun:2021aon}. 

\section{Matching for flavor-singlet quarks and gluons in non-forward kinematics}
\label{SEC:matchnonforward}
In the following, we calculate the matching kernel for non-forward qLFCs or for the GPDs. The calculation can be greatly simplified by observing that the four-momentum transfer $t=(P_1-P_2)^2$ only leads to power suppressed contributions, and thus do not contribute to the leading-twist matching. Therefore, we assume $t=0$ in the calculation hereafter. 
Another simplification comes from that the matching kernel is the same for the GPDs defined by the same nonlocal operator~\cite{Liu:2019urm}. Therefore, we do not need to consider all the GPDs for the calculation. 
Throughout the calculation, we use DR $(d=4-2\epsilon)$ to regularize both ultraviolet (UV) and infrared (IR) divergences. This also simplifies the calculation as scaleless integrals will vanish. 
We will first present the matching kernels in the $\overline{\rm MS}$ scheme, and then in the ratio and hybrid schemes.

\subsection{Matching kernel for quasi-light-front correlations}
The matching matrix for the non-forward qLFCs can be calculated by 
replacing the hadron states with quark or gluon parton states and expanding the factorization formula up to the NLO. The resulting expressions are summarized in Table~\ref{Tab:perexp},
\begin{table}[htbp]
  \centering
  \renewcommand\arraystretch{1.5}
  \begin{tabular}{ccc}
\hline
~~~ & $O_q/O_q^{l.t.}$ & $O_g/O_g^{l.t.}$ \\
\hline
\specialrule{0em}{2pt}{2pt}
quark & $
C_{qq}^{(1)}=\frac{\langle q| {O}_{q} |q' \rangle ^{(1)} -\langle  q| {O}_{q}^{l.t.} | q'\rangle ^{(1)} }{\langle q| {O}_{q}^{l.t.} | q' \rangle ^{(0)}}$ & $C_{gq}^{(1)}=\frac{\langle q| {O}_{g} | q' \rangle ^{(1)} -\langle  q| {O}_{g}^{l.t.} | q' \rangle ^{(1)} }{\langle q| {O}_{q}^{l.t.} |q' \rangle ^{(0)}}$ \\
\specialrule{0em}{2pt}{2pt}
gluon & $C_{qg}^{(1)}=\frac{\langle g| {O}_{q} |g' \rangle ^{(1)} -\langle  g| {O}_{q}^{l.t.} | g'\rangle ^{(1)} }{\langle g| {O}_{g}^{l.t.} |g' \rangle ^{(0)}}$ & $C_{gq}^{(1)}=\frac{\langle g| {O}_{g} |g' \rangle ^{(1)} -\langle  g| {O}_{g}^{l.t.} | g' \rangle ^{(1)} }{\langle g| {O}_{g}^{l.t.} |g' \rangle ^{(0)}}$ \\
\specialrule{0em}{2pt}{2pt}
\hline
\end{tabular}
\caption{The NLO matching matrix for non-forward qLFCs.}
 \label{Tab:perexp}
\end{table}
where $|q\rangle, |q'\rangle$ and $|g\rangle, |g'\rangle$ denote different on-shell quark and gluon external states 
with momenta $p_i^\mu=(p_i^0=p_i^z,0,0,p_i^z)$ with $i=1, 2$. The superscripts $(0)$ and $(1)$ denote the tree-level and NLO quantities, respectively. The factorization formula ensures that the IR divergence cancels out in the numerator of these expressions so that they are IR finite. 
{More generally, the formulas in the $\widebar{\rm MS}$ scheme can be written schematically as 
\begin{align}
    {O}_{i}^{R} = Z_{\rm UV}O_i^{B} = Z_{\rm UV}[C^B_{ij}\otimes O_j^B] = Z_{\rm UV} \left[C^B_{ij}Z^{-1}_{O_j}\otimes Z_{O_j}O_j^B\right] \equiv C_{ij}^R\otimes O^R_j\, ,
\end{align}
where $i,j\in {q,g}$ and the superscript $R$ ($B$) stands for renormalized (bare) quantities. 
The nonlocal {{(bare)}} collinear operators $O_i^{B}$ receive quantum corrections that are computable in perturbative QCD.
The multiplicative UV renormalization constants $Z_{\rm UV}$ remove the UV singularities and are known to three loops already~\cite{Braun:2020ymy}. The IR singularities, on the other hand, are removed by the renormalization constants of the specific LF operators.  Notice that we have inserted $ \mathbbm{1} = Z^{-1}_{O_j}\otimes Z_{O_j}$ to simultaneously remove the IR singularities present in both the bare coefficient function $C^B_{ij}$ and the LF operator $O^B_j$. 
The fact that the known renormalization constants $Z_{\rm UV}$ and $Z_{O_j}$ in, e.g.,~\cite{Belitsky:2005qn,Braun:2009vc,Ji:2014eta} completely remove all $1/\epsilon$ poles provide strong checks for our matching coefficients.
}

\subsubsection{$C_{qq}$}
The calculation of the matrix entry $C_{qq}$ is relatively simple, and has already been calculated in momentum space at one-loop in Refs.~\cite{Ji:2015qla,Xiong:2015nua,Liu:2019urm,Ma:2022ggj}. 
Here we give the results for the qLFCs, with the one-loop Feynman diagrams being shown in Fig.~\ref{fig:nonsin1loopns}.

\begin{figure}[htbp]
\centerline{\includegraphics[width=0.18\textwidth]{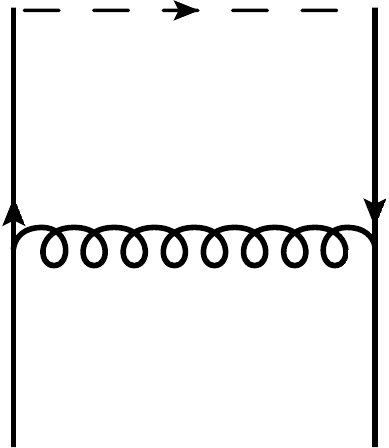}
\hspace*{3em}
\includegraphics[width=0.18\textwidth]{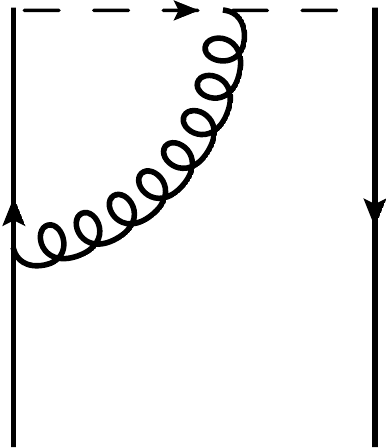}
\hspace*{3em}
\includegraphics[width=0.18\textwidth]{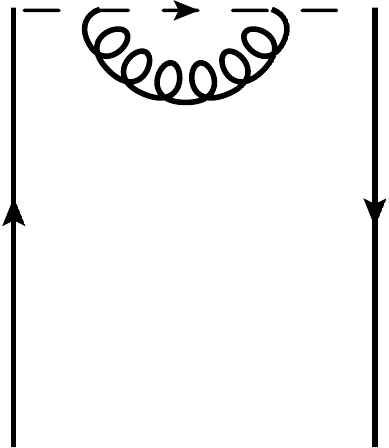}}
\caption{One-loop corrections to $C_{qq}$. The conjugate diagram to the second one and the quark self-energy diagram are not shown. }
\label{fig:nonsin1loopns}
\end{figure}

In the Feynman gauge, the ladder diagram yields the following contribution
\begin{align}
I_{1,qq}=&2a_s C_F(1-\epsilon)\Gamma(-\epsilon)\,{ \eta}^{\epsilon}\, A_1
\int_0^1 d\alpha\int_0^{\bar\alpha}d\beta\, O_q(z_{12}^\alpha,z_{21}^\beta)\, , \notag\\
	&A_{1,\,u}=1\,, \qquad A_{1,\,h}=1-2\epsilon\, , \qquad A_{1,\,t}=0\, ,\notag
\end{align} 
where $a_s=\alpha_s/(4\pi)$
and $\eta=-(z_{12}^2\mu^2 e^{\gamma_E})/4$ in the $\widebar{\rm MS}$ scheme. The calculation is done 
with the collinear on-shell condition of the quark momenta $p_1^2=p_2^2=0$, $p_1 \parallelsum p_2$\footnote{More generally, $p_1$ and $p_2$ are understood to be the parton momenta, {and are related to the hadron momentum as $p_i^z=y_i P_i^z$. Here we use $p_i$ as a short-hand notation in intermediate steps.}}. Note that only the ladder diagram depends on the Dirac structure in the quark operator. The contribution of the vertex diagram and its conjugate are independent of the Dirac structure, and takes the following form
\begin{align}\label{Eq:C_qq_qLFCs_2}
I_{2,qq}=-2a_s C_F\Gamma(-\epsilon)\eta^\epsilon
 \int_0^1 \frac{d\alpha}{\alpha}&\bigg[
 \frac{1-\epsilon}{1-2\epsilon}(\alpha-\alpha^{2\epsilon})
\left(O_q(z_{12}^\alpha,z_2)+O_q(z_1, z_{21}^\alpha)\right) \notag\\ 
&+2\alpha^{2\epsilon}O_q(z_1,z_2)\bigg].
\end{align}
The Wilson line self-energy diagram yields
\begin{align}\label{Eq:C_qq_qLFCs_3}
I_{3,qq}&=-\frac{2a_s C_F\Gamma(-\epsilon)\,\eta^\epsilon}{1-2\epsilon}O_q(z_1,z_2)\, .
\end{align}
Since we do not distinguish UV and IR divergences, the quark self-energy diagram vanishes in DR. Also the one-loop matrix element of $O_q^{l.t.}$ gives zero. Adding up all the contributions, we obtain the one-loop matching kernel for the quark qLFCs
in the $\overline{\rm MS}$ scheme
\begin{align}\label{eq:coo_C_qq}
   C_{qq}^{\overline{\rm{MS}}}(\alpha,\beta,\mu^2 z_{12}^2)&=\delta(\alpha)\delta(\beta)+2a_sC_F \bigg\{ \left(A_{2} + \left[\frac{\bar{\alpha}}{\alpha}\right]_+\delta(\beta) + \left[\frac{\bar{\beta}}{\beta}\right]_+ \delta(\alpha) \right)(\rm{L_z}-1)+A_{3} \,\notag\\
	&-2\left[\frac{\ln(\alpha)}{\alpha}\right]_+\delta({\beta}) -2\left[\frac{\ln(\beta)}{\beta}\right]_+\delta({\alpha})   \bigg\}+2a_sC_F (-2\rm{L_z}+2)\delta(\alpha)\delta(\beta)\,, \notag\\
        &~~~~~~~~A_{2,\,u}=1\,, \qquad A_{2,\,h}=1\, , \qquad 
        A_{2,\,t}=0\, ,\notag\\
        &~~~~~~~~A_{3,\,u}=2\,, \qquad A_{3,\,h}=4\, , \qquad A_{3,\,t}=0\, ,
\end{align}
where ${\rm{L_z}}=\ln\frac{4 e^{-2\gamma_E}}{-\mu^2 z_{12}^2}$. {Throughout this work, we set $\mu_{ R} = \mu_F=\mu$ for convenience with $\mu_R$ and $\mu_{\rm F}$ being the renormalization and factorization scale, respectively.} The ``+'' subscript denotes the usual plus-prescription for the singularity 
\begin{align}
	\left[\frac{f(\alpha)}{\alpha}\right]_+= \frac{f(\alpha)}{\alpha}-\delta(\alpha)\int^1_0 \frac{f(\alpha')}{\alpha'}d\alpha'.
\end{align}

\subsubsection{$C_{qg}$}
Next we consider $C_{qg}$, which requires calculating the gluon matrix element of the quark operator, and is non-zero only for the unpolarized and helicity operators. The corresponding diagram is shown in Fig.~\ref{diag:qg1}.
\begin{figure}[!ht]
\begin{center}
\includegraphics[width=0.25\textwidth]{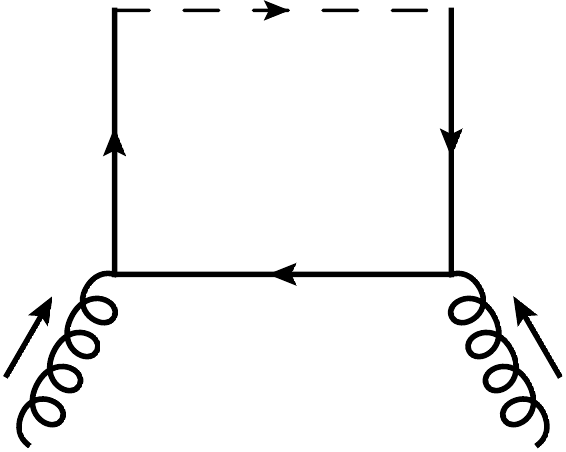}
\end{center}
\caption{One-loop correction to $C_{qg}$. The cross diagram is not shown.}
\label{diag:qg1}
\end{figure}

This diagram gives
\begin{align}\label{eq:Iqg}
I_{1,qg}&=2a_sT_F N_f\Gamma(-\epsilon) \frac{\eta^\epsilon}{{\mathbf z}_{12}} \hat{T} A_\rho(p_1)\cdot A_\sigma(p_2) \int_0^1 d\alpha\int_0^{\bar\alpha}d\beta\left\{B_1\, (p_1\cdot z_{12})-B_2\, (p_2\cdot z_{12}) \right.\notag\\
& \left.\quad +i\alpha\bar\alpha (p_1\cdot z_{12})^2 +i\beta\bar\beta (p_2\cdot z_{12})^2 +i (\bar\alpha\bar\beta+\alpha\beta)(p_1\cdot z_{12})(p_2\cdot z_{12})\right\}\e^{-i({p_1\cdot z}_{12}^{\alpha} +p_2\cdot{z}_{21}^\beta )}\, ,\notag\\
B_{1,u}&=\bar\alpha \,,\qquad B_{1,h}=1-3\alpha \,,\qquad
B_{2,u}=\bar\beta\, ,\qquad
B_{2,h}=1-3\beta\, ,
\end{align}
where $\hat{T}=\{g_{\perp}^{\rho\sigma},\epsilon_{\perp}^{\rho\sigma}\}$ correspond to the unpolarized and helicity case,
respectively. We have applied $z_{12}\cdot A =p_1\cdot A=p_2\cdot A=0$ as our external gluons are physical on-shell states. {{Since 
gauge invariance guarantees that the matching function is convoluted with the whole $O_g$ operator, it is sufficient to extract the coefficient function that convolutes with $z_{12}^\alpha\partial_\alpha A_\mu$ which is part of $F_{z_{12}\mu}$}}. In the lightcone limit, we have $F_{z_{12}\mu}\propto F_{+\mu}$ recovering the LF gluon GPD.
The cross diagram $I_{2,qg}$ can be obtained from $I_{1,qg}$ by interchanging $z_1\leftrightarrow z_2$ and multiplying with an overall factor $-1$. Taking into account that gluons are bosons and making  change of variables, 
we find
\begin{align}
I_{2,qg}=I_{1,qg}.
\end{align}
Thus, we get the final expression of the qLFCs in the $\widebar{\rm MS}$ scheme
\begin{align}\label{eq:Ccalqg}
I_{qg}&=2 I_{1,qg}=4ia_sT_FN_f\Gamma(-\epsilon) {\eta^\epsilon}\,{{\mathbf z}_{12}} \int^1_0d\alpha\int^{\bar\alpha}_0 d\beta\,B_{3} \,{O}_g(z_{12}^\alpha, z_{21}^\beta)\, ,\notag\\
B_{3,u}&=\bar\alpha\bar\beta+3\alpha\beta \, ,\qquad
B_{3,h}=\bar\alpha\bar\beta-\alpha\beta .
\end{align}
Putting everything together, we are able to write down the matching kernel as
\begin{align}
 C_{qg}^{\overline{\rm{MS}}}(\alpha,\beta,\mu^2 z_{12}^2)=4ia_sT_F N_f \,{\mathbf z_{12}} \,B_3\, {\rm{L_z}} \, . 
\end{align}

\subsubsection{$C_{gq}$}
The calculation of $C_{gq}$ requires calculating the quark matrix element of the gluon operator, and the corresponding diagram is shown in Fig.~\ref{diag:gq1}. 
\begin{figure}[h]
\begin{center}
\includegraphics[width=0.25\textwidth]{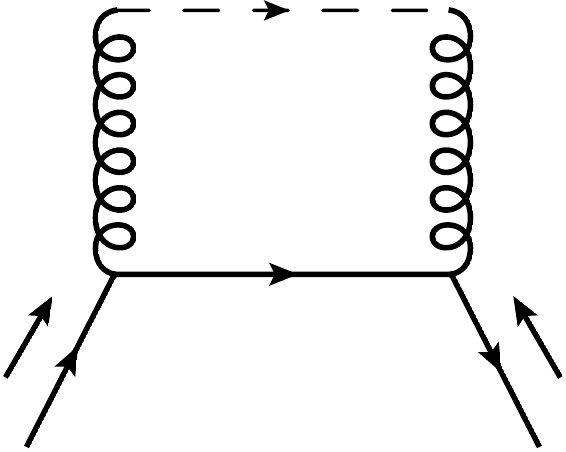}
\end{center}
\caption{One-loop correction to $C_{gq}$. The cross diagram is not shown.}
\label{diag:gq1}
\end{figure} 

The Feynman diagram yields
\begin{align}
I_{1,gq}^u=&\frac{-ia_sC_F\Gamma(-\epsilon)\eta^\epsilon}{\mathbf z_{12}} \bar q(p_1)\Gamma q(p_2) \int^1_0d\alpha\int^{\bar\alpha}_0 d\beta\,\e^{-i(z_{12}^\alpha\cdot p_1+z_{21}^\beta\cdot p_2)}(1-\epsilon)\notag\\
& \bigg\{ \bigg((p_1\cdot z_{12})^2 \alpha\bar\alpha+2i(p_1\cdot z_{12})(2\epsilon \alpha- \bar\alpha-\epsilon)+(\alpha \leftrightarrow \beta, p_1 \leftrightarrow -p_2)\bigg)\notag\\
&+(p_1\cdot z_{12})(p_2\cdot z_{12})(\alpha\beta+\bar\alpha\bar\beta)+4\epsilon (1+\epsilon)\bigg\}\, ,\notag\\
I_{1,gq}^h=&\frac{-ia_sC_F\Gamma(-\epsilon)\eta^\epsilon}{\mathbf z_{12}} \bar q(p_1)\Gamma q(p_2) \int^1_0d\alpha\int^{\bar\alpha}_0 d\beta\,\e^{-i(z_{12}^\alpha\cdot p_1+z_{21}^\beta\cdot p_2)} \bigg\{2\epsilon(2\epsilon+3)\notag\\
& + \bigg((p_1\cdot z_{12})^2 \alpha\bar\alpha+2i(p_1\cdot z_{12})(1+\epsilon)(2\alpha-1)+(\alpha \leftrightarrow \beta, p_1 \leftrightarrow -p_2)\bigg)\notag\\
&+(p_1\cdot z_{12})(p_2\cdot z_{12})(\alpha\beta+\bar\alpha\bar\beta)\bigg\}\, ,
\end{align}
where $\Gamma=\{\gamma^z,\gamma^z \gamma_5\}$ denotes the generated structure in the unpolarized and helicity case, respectively.
The cross diagram $I_{2,gq}$ is obtained by $z_1 \leftrightarrow z_2$ and multiplying with an overall factor $-1$. Adding them up, the total contribution reads
\begin{align}
I_{gq}&= I_{1,gq}+I_{2,gq}\notag\\
&= \frac{-2ia_sC_F\Gamma(-\epsilon)\eta^\epsilon}{{\mathbf z_{12}}}\,\bigg\{ D_1\, O_q^s(z_1,z_2) +2 \epsilon \, D_1 \int^1_0d\alpha \left[O_q^s(z_1,z_{21}^\alpha)+O_q^s(z_{12}^\alpha,z_2)\right]\notag\\
&~~~~~~~~~~~~~~~~~~~~~~~~~~~+ 2\,D_2\int^1_0d\alpha\int^{\bar\alpha}_0 d\beta\, O_q^s(z_{12}^\alpha,z_{21}^\beta)\bigg\}\, ,\notag\\
D_{1,u}&=1-\epsilon\, ,  \quad D_{1,h}=1\,  ,\quad 
D_{2,u}=1-3\epsilon+4\epsilon^2-2\epsilon^3\, ,  \quad D_{2,h}=2\epsilon^2-\epsilon-1\,  .
\end{align}
From the above result, we obtain the matching kernel in the $\widebar{\rm MS}$ scheme as
\begin{align}\label{eq:Cgq-pos}
  C_{gq}^{\overline{\rm{MS}}}(\alpha,\beta,\mu^2 z_{12}^2) &= \frac{-2ia_sC_F}{{\mathbf z_{12}}}\bigg\{ D_{3} \left({\rm{L_z}}+1\right)+4-2\big(\delta(\alpha)+\delta(\beta)\big)+\left({\rm{L_z}}+D_4\right) \delta(\alpha)\delta(\beta) \bigg\}\,,\notag\\
	D_{3,u}&=2\,, \qquad D_{3,h}=-2\, , \qquad D_{4,u}=1\, ,  \qquad D_{4,h}=0\,  .
\end{align}

\subsubsection{{$C_{gg}$}}
The calculation of $C_{gg}$ requires calculating the gluon matrix element of the gluon operator. The corresponding Feynman diagrams are depicted in Fig.~\ref{diag:gg-1}. 
\begin{figure}[h]
\begin{center}
		\begin{overpic}[width=0.8\textwidth]{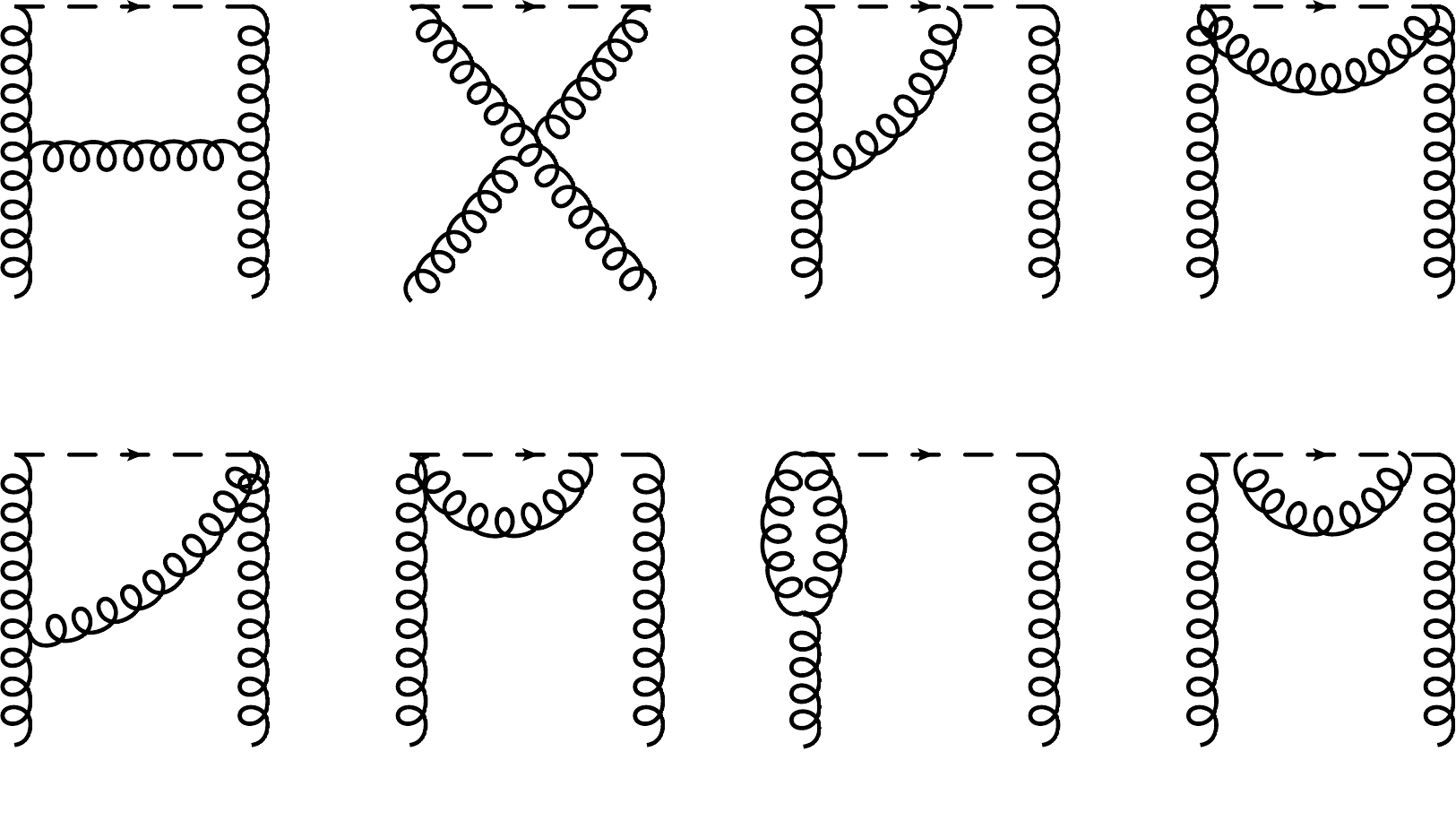}
	\put(9,32){\small$\text{I}_1$}
        \put(36,32){\small$\text{I}_2$}
        \put(63,32){\small$\text{I}_3$}
        \put(90.5,32){\small$\text{I}_4$}
	\put(9,1){\small$\text{I}_5$}
        \put(36,1){\small$\text{I}_6$}
        \put(63,1){\small$\text{I}_7$}
        \put(90.5,1){\small$\text{I}_8$}
		\end{overpic}
\end{center}
\caption{One-loop corrections to $C_{gg}$. 
The conjugate diagrams and self-energy diagrams are not shown. $I_2$ represents the contribution of four-gluon vertex.}
\label{diag:gg-1}
\end{figure}
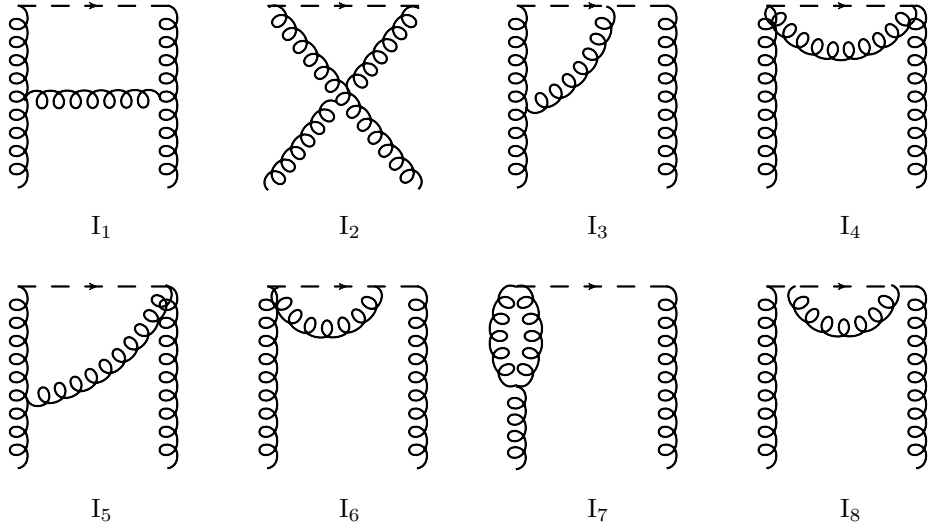

The ladder diagram gives
\begin{align}
I_{1,gg}^u&=\frac{a_sC_A\Gamma(-\epsilon)\eta^\epsilon}{{\mathbf z}_{12}^2}\hat{T} A_\rho(p_1) A_\sigma(p_2)\int^1_0d\alpha\int^{\bar\alpha}_0 d\beta\,\e^{-i(z_{12}^\alpha\cdot p_1+z_{21}^\beta\cdot p_2)}\bigg\{4\epsilon(1-4\epsilon^2+6\epsilon)\notag\\
&\quad-\bigg(2i\epsilon\,(7-2\epsilon+\alpha(6\epsilon-13))(p_1\cdot z_{12})+i\bar\alpha[\bar\beta-\alpha(2\bar\beta-\beta)](p_1\cdot z_{12})^2(p_2\cdot z_{12})\notag\\
&\quad+(3-\alpha-8\alpha\bar{\alpha}+2\epsilon\bar{\alpha})(p_1\cdot z_{12})^2+i\alpha\bar\alpha^2(p_1\cdot z_{12})^3+(\alpha \leftrightarrow \beta, p_1\leftrightarrow -p_2 )\bigg)\notag\\
&\quad+(9\bar\alpha\bar\beta+7\alpha\beta-1-2\epsilon(2+\alpha+\beta))(p_1\cdot z_{12})(p_2\cdot z_{12})\bigg\},\notag\\
I_{1,gg}^h&=\frac{a_sC_A\Gamma(-\epsilon)\eta^\epsilon}{{\mathbf z}_{12}^2} \hat{T} A_\rho(p_1) A_\sigma(p_2)\int^1_0d\alpha\int^{\bar\alpha}_0 d\beta\,\e^{-i(z_{12}^\alpha\cdot p_1+z_{21}^\beta\cdot p_2)}\bigg\{4\epsilon(2\epsilon-1)\notag\\
&\quad-\bigg(2i\epsilon\,(2\bar{\alpha}\epsilon+7-5\alpha)(p_1\cdot z_{12})+(4\alpha^2(1+\epsilon)+(3+2\epsilon)(1-3\alpha))(p_1\cdot z_{12})^2\notag\\
&\quad+i\bar\alpha[\bar\beta-\alpha(2\bar\beta-\beta)](p_1\cdot z_{12})^2(p_2\cdot z_{12})+i\alpha\bar\alpha^2(p_1\cdot z_{12})^3+(\alpha \leftrightarrow \beta, p_1\leftrightarrow -p_2 )\bigg)\notag\\
&\quad+(\alpha(8\beta(\epsilon+1)-6\epsilon-9)-6\beta\epsilon+4\epsilon+8)(p_1\cdot z_{12})(p_2\cdot z_{12})\bigg\},\notag\\
I_{1,gg}^t&=\frac{a_sC_A\Gamma(-\epsilon)\eta^\epsilon}{{\mathbf z}_{12}^2} \hat{S} A_\rho(p_1) A_\sigma(p_2)\int^1_0d\alpha\int^{\bar\alpha}_0 d\beta\,\e^{-i(z_{12}^\alpha\cdot p_1+z_{21}^\beta\cdot p_2)}\bigg\{4\epsilon(2\epsilon+1)\notag\\
&\quad-\bigg(2i\epsilon\,(2\bar{\alpha}\epsilon+3-5\alpha)(p_1\cdot z_{12})+(4\alpha^2(1+\epsilon)-(6\epsilon+5)\alpha+2\epsilon+3)(p_1\cdot z_{12})^2\notag\\
&\quad+i\bar\alpha[\bar\beta-\alpha(2\bar\beta-\beta)](p_1\cdot z_{12})^2(p_2\cdot z_{12})+i\alpha\bar\alpha^2(p_1\cdot z_{12})^3+(\alpha \leftrightarrow \beta, p_1\leftrightarrow -p_2 )\bigg)\notag\\
&\quad+(\alpha(8\beta(\epsilon+1)-6\epsilon-5)-\beta(6\epsilon+5)+4\epsilon)(p_1\cdot z_{12})(p_2\cdot z_{12})\bigg\},
\end{align}
where $\hat{T}$ is the same as before for the unpolarized and helicity operators, and $\hat{S}$ has been defined in Eq.~(\ref{eq:gluonope}). 
The second diagram is
\begin{align}
I_{2,gg}^u&=-\frac{2a_sC_A\Gamma(-\epsilon)\eta^\epsilon}{{\mathbf z}_{12}^2}\hat{T} A_\rho(p_1) A_\sigma(p_2)\int^1_0d\alpha\,\e^{-i(p_1+p_2)\cdot z_{12}^\alpha}\,\bigg\{(1-2\epsilon)\alpha\bar\alpha [(p_1+p_2)\cdot z_{12}]^2\notag\\
&\quad-2i\epsilon(1-\epsilon)(1-2\alpha)(p_1+p_2)\cdot z_{12}+2\epsilon(1+2\epsilon-4\epsilon^2)\bigg\}\, ,\notag\\
I_{2,gg}^h&=0, \notag\\
I_{2,gg}^t&=-\frac{2a_sC_A\Gamma(-\epsilon)\eta^\epsilon}{{\mathbf z}_{12}^2}\hat{S} A_\rho(p_1) A_\sigma(p_2)\int^1_0d\alpha\,\e^{-i(p_1+p_2)\cdot z_{12}^\alpha}\,\bigg\{-\alpha\bar\alpha [(p_1+p_2)\cdot z_{12}]^2\notag\\
&\quad+2i\epsilon(1-2\alpha)(p_1+p_2)\cdot z_{12}+2\epsilon(1-2\epsilon)\bigg\}\, .
\end{align}
These two diagrams are the only diagrams that depend on the choice of operators. The other diagrams yield the same result for unpolarized, helicity and transversity operators apart from the generated tree-level structure (indicated by $\{\hat{T},\hat{S}\}$ below)
\begin{align}
I_{3,gg}=&-\frac{a_sC_A\Gamma(-\epsilon)\eta^\epsilon}{{\mathbf z}_{12}^2}\,\{\hat{T},\hat{S}\}A_\rho(p_1)  F_{\sigma z_{12}}(z_2)\int^1_0\frac{du}{\bar u^{2-2\epsilon}} \int^1_0d\alpha\,\e^{-ip_1\cdot(z_1-\bar u\alpha z_{12})}\bigg\{2\epsilon(1+2\epsilon)\notag\\
&\qqquad+2i\epsilon\bar u(2\alpha-3)(p_1\cdot z_{12})-\bar u^2\bar\alpha(2-\alpha)(p_1\cdot z_{12})^2\bigg\}\, .
\end{align}
$I_{3,gg}$ contains the conjugate diagram which can be obtained by making the replacement $z_1\leftrightarrow z_2$, $p_1\leftrightarrow p_2$ and $\alpha \rightarrow 1-\alpha $ in the contribution of the third diagram. In addition, $F_{\sigma z_{12}}(z_2)$ can be expanded as following
\beq
F_{\sigma z_{12}}(z_2)= i(p_2\cdot z_{12})A_\sigma (p_2) e^{-i p_2\cdot z_2}.
\eeq
The next three diagrams give
\begin{align}
I_{4,gg}=&-\frac{4a_sC_A\Gamma(1-\epsilon)\eta^\epsilon}{{\mathbf z}_{12}^2}\,\{\hat{T},\hat S\}A_\rho(p_1) A_\sigma(p_2)e^{-i(p_1\cdot z_1+p_2\cdot z_2)}\, ,\notag\\
I_{5,gg}=& -\frac{a_sC_A\Gamma(-\epsilon)\eta^\epsilon}{{\mathbf z}_{12}^2}\,\{\hat{T},\hat S\}A_\rho(p_1) A_\sigma(p_2)\int^1_0d\alpha\,\e^{-ip_1\cdot z_{12}^\alpha} \bigg\{-2i(3-2\alpha)\epsilon(p_1\cdot z_{12})\notag\\
&-\bar\alpha(2-\alpha)(p_1\cdot z_{12})^2+2\epsilon(1+2\epsilon)\bigg\}\,,\notag\\
I_{6,gg}=&-\frac{4a_s^2C_A\Gamma(1-\epsilon) \eta^\epsilon}{ (2\epsilon-1){\mathbf z}_{12}^2}\,\{\hat{T},\hat S\}A_\rho(p_1) F_{\sigma z_{12}}(z_2)e^{-ip_1 \cdot z_1}\, .
\end{align}
To proceed, we need to recast $I_{4,gg}$, $I_{6,gg}$ into a form dictated by the factorization theorem, or in other words, to a form similar to $I_{1,gg}$. This can be done by using some integration tricks. For example, the term $A_\rho(p_1) A_\sigma(p_2)$ can be written as the form of two Feynman variables $\alpha$ and $\beta$,
\begin{align}
& A_\rho(p_1) A_\sigma(p_2)e^{-i(p_1\cdot z_1+p_2\cdot z_2)}\notag\\
=& A_\rho(p_1) A_\sigma(p_2)\int^1_0d\alpha\, \frac{d}{d\alpha} \int^{\bar\alpha}_0d\beta\,\frac{d}{d\beta}\,(1-\alpha-\beta) \e^{-i(p_1\cdot z_{12}^\alpha+p_2\cdot z_{21}^\beta)}\notag\\
=&A_\rho(p_1) A_\sigma(p_2)\int^1_0d\alpha\bigg[\left(1-i(\bar\alpha p_1-\alpha p_2)\cdot z_{12}\right)\e^{-i(p_1+p_2)\cdot z_{12}^\alpha}+ (p_1\cdot z_{12})(p_2\cdot z_{12})\notag\\
& ~~~~~~~~~~~~~~~~~~~~~~~~\times\int^{\bar\alpha}_0d\beta \e^{-i(p_1\cdot z_{12}^\alpha+p_2\cdot z_{21}^\beta)}\bigg]\, .
\end{align}
$I_7$ contains scaleless integrals only, as can be seen after writing down the explicit amplitude
\begin{align}
I_{7,gg}&=\frac{g^2C_A}{{\mathbf z}_{12}^2}z_{12}^\mu \hat{T} A_\rho(p_1)F_{\sigma z_{12}}(z_2)\int\frac{d^dl}{(2\pi)^d}\,\frac{g_{\alpha\rho}(l-2p_1)_\beta+g_{\rho\beta}(l+p_1)_\alpha+g_{\alpha\beta}(p_1-2l)_\rho}{l^2(l-p_1)^2}\notag\\
& \quad \times \Big(g_{\mu\alpha}g_{\nu\beta} -g_{\mu\beta}g_{\nu\alpha} \Big) e^{-ip_1\cdot z_1}\notag\\
&=0 .
\end{align}
Thus, it vanishes in DR. 
The Wilson-line self-energy can be directly written down,
\begin{align}
I_{8,gg}&=-\frac{2a_sC_A\Gamma(-\epsilon) \eta^\epsilon}{(1-2\epsilon){\bf z}_{12}^2}\,{{\{\hat{T},\hat S\}}} F_{z_{12} \rho}(z_1)F^{\sigma z_{12}}(z_2)\, .
\end{align}
Adding all contributions together, including the conjugate diagrams, and using integration-by-parts (IBP) techniques and boson symmetry (symmetry under $p_1\leftrightarrow p_2$ and $z_1\leftrightarrow z_2$),
we find that all structures with the phase factor $e^{-i(p_1+p_2)\cdot z_{12}^\alpha}$ cancel out, as required by the factorization theorem. 
Now we are able to extract the total one-loop coefficient function, which reads
\begin{align}
 C_{gg}^{\overline{\rm{MS}}}(\alpha,\beta,\mu^2 z_{12}^2) =& 
 \delta(\alpha)\delta(\beta)
 +2a_sC_A \bigg\{ \left(E_{1} + \left[\frac{\bar{\alpha}^2}{\alpha}\right]_+\delta(\beta) + \left[\frac{\bar{\beta}^2}{\beta}\right]_+ \delta(\alpha) \right)\left({\rm{L_z}}-1\right)+E_{2}\notag\\
 &-2\left[\frac{\ln(\alpha)}{\alpha} \right]_+\delta({\beta})-2\left[\frac{\ln(\beta)}{\beta}\right]_+\delta({\alpha})   \bigg\}+2a_sC_A \left(-3\,{\rm{L_z}}+2\right)\delta(\alpha)\delta(\beta)\,, \notag
 \end{align}
\begin{align}
E_{1,u}=&4(1-\alpha-\beta+3\alpha\beta)\, , \qquad  E_{1,h}=4(1-\alpha-\beta)\, , \qquad  E_{1,t}=0 ,\notag\\
E_{2,u}=&\frac{5}{2}\,E_{1,u}+6\alpha\beta\, , \qquad  E_{2,h}=\frac{3}{2}\,E_{1,h}\, , \qquad  E_{2,t}=2(1+\alpha+\beta-2\alpha\beta) .
\end{align}

{The above calculations present the complete one-loop
matching kernels in coordinate space that appear in the factorization of quark and gluon quasi-LF correlations in non-forward kinematics. In Ref.~\cite{Radyushkin:2019owq}, the authors provide the one-loop correction for $C_{qq,u}$ which serves as a useful check for our result.}

\subsection{Pseudo-GPDs and quasi-GPDs}
In this subsection, we present the matching {{kernels}} for the pGPDs and qGPDs.

\subsubsection{Matching kernel in pseudo space}
Following the discussion in previous subsections, the one-loop matching kernels of the pGPDs can be obtained from that in coordinate space by a Fourier transform in Eq.~(\ref{eq:pseudoFT}), and take the following form in the $\overline{\rm MS}$ scheme: 
\begin{itemize}
    \item Quark in quark
\end{itemize}
\begin{align}\label{eq:ms_exp_pseudo}
	{\cal{C}}_{qq}^{\overline{\rm{MS}}}(\tau_1,\tau_2,y_1,y_2;\mu^2 z_{12}^2)
	=&\delta(\tau_1-y_1)+a_s C_F {\cal{C}}^{(1)}_{qq}(\tau_1,\tau_2,y_1,y_2;\mu^2 z_{12}^2)
\end{align}
with
\begin{align}\label{eq:pseudoPqq}
	{\cal{C}}^{(1)}_{qq,u}
	&=\bigg\{ \frac{ |\tau_1| }{y_1 (\tau_1+\tau_2)} +\frac{|\tau_2|}{y_2(\tau_1+\tau_2)}-\frac{|\tau_1-y_1|}{y_1y_2} \bigg\}\left( {\rm{L_z}}+1 \right)+{\cal{C}}^{(1)}_{qq,t} \,  ,\notag\\
	{\cal{C}}^{(1)}_{qq,h}
	&={\cal{C}}^{(1)}_{qq,u}+2\, \bigg\{ \frac{ |\tau_1| }{y_1 (\tau_1+\tau_2)} +\frac{|\tau_2|}{y_2(\tau_1+\tau_2)}-\frac{|\tau_1-y_1|}{y_1y_2}  \bigg\} \,  ,\notag\\	
	{\cal{C}}^{(1)}_{qq,t}
	&=\bigg\{ \left(\frac{|\tau_1|}{y_1(y_1-\tau_1)}+\frac{\tau_1}{y_1|\tau_1-y_1|} \right)\left( {\rm{L_z}}-1 \right)  
	+\left( \frac{|\tau_1|}{\tau_1(\tau_1-y_1)}-\frac{1}{|\tau_1-y_1|} \right) \ln\frac{(\tau_1-y_1)^2}{y_1^2}\notag\\
	&~~~~~+(\tau_1 \rightarrow \tau_2,\,y_1 \rightarrow y_2) \bigg\} _+ + 2\, \left(-2\,{\rm{L_z}}+2\right) \delta(\tau_1-y_1) \, .
\end{align}
The singularity as $y_i\rightarrow x_i$ gets regulated by the plus-prescription
\begin{align}
	[f(x_i,y_i)]_+= f(x_i,y_i)-\delta(x_i-y_i)\int f(x_i',y_i')\,dx_i'\, ,
\end{align}
where the integral limits are determined by the range of the momentum fractions in the LF distributions. 

Note that the Fourier transform on the $C_{gq}$ channel~\eqref{eq:Cgq-pos} can be most easily done by writing
\begin{align}
\frac{1}{\bf z_{12}} \e^{ix{\bf z_{12}}}=i\int^{x}_{\infty(-1+i\epsilon)}\e^{ix_1{\bf z_{12}}}\, dx_1 \, ,\qquad {\text{assuming} \quad {\bf z}_{12}>0}\, ,
\end{align}
where the lower integral limit ${\infty(-1+i\epsilon)}$ is chosen for convenience. Such a choice is, however, arbitrary and therefore brings ambiguities.  
The ambiguities are removed by requiring that the matching coefficients in the pseudo and coordinate space must generate identical Mellin moments (for more details, see discussions in Appendix~\ref{sec:app1}). 

In the following, we list the results for ${\cal{C}}_{qg}$, ${\cal{C}}_{gq}$ and ${\cal{C}}_{gg}$:
\begin{itemize}
    \item Quark in gluon
\end{itemize}
\begin{align}\label{}
	{\cal{C}}_{qg}^{\overline{\rm{MS}}}(\tau_1,\tau_2,y_1,y_2;\mu^2 z_{12}^2)
	=&{-}a_sT_FN_f \, {\cal{C}}^{(1)}_{qg}(\tau_1,\tau_2,y_1,y_2;\mu^2 z_{12}^2)
\end{align}
with
\begin{align}\label{}
    {\cal{C}}_{qg,u}^{(1)}
    &=\, \bigg\{\frac{|\tau_1|(2\tau_2-y_2)}{y_1^2y_2(y_1+y_2)} +\frac{(\tau_2-y_2)^2-\tau_1 \tau_2}{y_1^2y_2^2(\tau_1-y_1)} |\tau_1-y_1|-\frac{|\tau_2|(2\tau_1-y_1)}{y_1y_2^2(y_1+y_2)}\bigg\}\,{\rm{L_z}}\,,\notag\\
	{\cal{C}}_{qg,h}^{(1)}
	&=\, \bigg\{ \frac{|\tau_1|}{(\tau_1+\tau_2)y_1^2} +\frac{\tau_1(y_1-y_2)-y_1^2}{y_1^2y_2^2(\tau_1-y_1)} |\tau_1-y_1|-
	\frac{|\tau_2|}{(\tau_1+\tau_2)y_2^2}\bigg\}\,{\rm{L_z}}\, .
\end{align}

\begin{itemize}
    \item Gluon in quark
\end{itemize}
\begin{align}\label{}
	{\cal{C}}_{gq}^{\overline{\rm{MS}}}(\tau_1,\tau_2,y_1,y_2;\mu^2 z_{12}^2)
	=&-2a_sC_F\,{\cal{C}}^{(1)}_{gq}(\tau_1,\tau_2,y_1,y_2;\mu^2 z_{12}^2)
\end{align}
with
\begin{align}\label{}
	{\cal{C}}_{gq,u}^{(1)}
	=&\,\bigg\{  \left(\frac{|\tau_1|(\tau_1-2 y_1)}{y_1(y_1+y_2)} -\frac{(\tau_1-y_1)^2-y_1 y_2}{y_1 y_2(\tau_1-y_1)} |\tau_1-y_1|-\frac{|\tau_2|(\tau_2-2 y_2)}{y_2(y_1+y_2)}\right) \left({\rm{L_z}+1}\right) \notag\\
	&~~~+\frac{2|\tau_1|(\tau_1-y_2)}{y_1(y_1+y_2)} -\frac{2|\tau_1-y_1|(\tau_1-y_2)}{y_1 y_2}-\frac{2|\tau_2|(\tau_2-y_1)}{y_2(y_1+y_2)}\bigg\}\,,\notag\\
	{\cal{C}}_{gq,h}^{(1)}
	=&\,\bigg\{ \left(-\frac{|\tau_1|\tau_1}{y_1(y_1+y_2)} +\frac{(\tau_1-y_1)^2+y_1 y_2}{y_1 y_2(\tau_1-y_1)} |\tau_1-y_1|+\frac{|\tau_2|\tau_2}{y_2(y_1+y_2)}\right) \left({\rm{L_z}-1}\right) \notag\\
	&~~~+\frac{|\tau_1|(y_1-2 y_2)}{y_1(y_1+y_2)} +\frac{y_1(2y_1-y_2)+2\tau_1(y_2-y_1)}{y_1 y_2(\tau_1-y_1)} |\tau_1-y_1|-\frac{|\tau_2|(y_2-2 y_1)}{y_2(y_1+y_2)}\bigg\}\, .	
\end{align}
\begin{itemize}
    \item Gluon in gluon
\end{itemize}
\begin{align}\label{}
	{\cal{C}}_{gg}^{\overline{\rm{MS}}}(\tau_1,\tau_2,y_1,y_2;\mu^2 z_{12}^2)
	=&\delta(\tau_1-y_1)+a_sC_A\,{\cal{C}}^{(1)}_{gg}(\tau_1,\tau_2,y_1,y_2;\mu^2 z_{12}^2)
\end{align}
with
\begin{align}\label{}
	{\cal{C}}_{gg,u}^{(1)}
	&= \bigg\{ \frac{ 2|\tau_1|\tau_1((y_1+y_2)(4y_1+y_2)-\tau_1(3y_1+y_2))}{ y_1^2 (y_1+y_2)^3}\left({\rm{L_z}}+2\right)-\frac{|\tau_1|\tau_1}{y_1^2(y_1+y_2)}\notag\\
	&~~~~-\frac{|\tau_1-y_1|(\tau_1\tau_2+y_1y_2)}{y_1^2y_2^2} \left({\rm{L_z}}+\frac{3}{2}\right)+\frac{|(\tau_1-y_1)^3|}{2y_1^2y_2^2}+(\tau_1 \leftrightarrow \tau_2,\,y_1 \leftrightarrow y_2)\bigg\}+{\cal{T}}_{gg}^{(1)} \,  ,\notag\\
	{\cal{C}}_{gg,h}^{(1)}
	&=\bigg\{ \frac{ 2|\tau_1|\tau_1 }{y_1^2 (y_1+y_2)} +\frac{2|\tau_2|\tau_2}{y_2^2(y_1+y_2)}-\frac{2|\tau_1-y_1|(\tau_1 y_2+\tau_2 y_1)}{y_1^2y_2^2} \bigg\}\left({\rm{L_z}}+\frac{1}{2}\right)+{\cal{T}}_{gg}^{(1)} \,  ,\notag\\	
	{\cal{C}}_{gg,t}^{(1)}
	&=\bigg\{  \left(\frac{4 |\tau_1|}{y_1(y_1+y_2)}-\frac{|\tau_1|\tau_1 (\tau_1+3 \tau_2) (3y_1+y_2)}{3 {y_1}^2 (y_1+y_2)^3} + (\tau_1 \leftrightarrow \tau_2, \,y_1 \rightarrow y_2) \right)\notag\\
	&~~~~+\frac{|\tau_1-y_1|(-2\tau_1^2+y_1(y_1-9y_2)+\tau_1(y_1+3y_2))}{3y_1^2y_2^2} \bigg\} +{\cal{T}}_{gg}^{(1)} \, , \notag\\
	{\cal{T}}_{gg}^{(1)}
	&=\bigg\{ \left(\frac{\tau_1|\tau_1|}{y_1^2(y_1-\tau_1)}+\frac{\tau_1^2}{y_1^2|\tau_1-y_1|} \right)({\rm{L_z}}-1)  
	+\left(\frac{|\tau_1|}{\tau_1(\tau_1-y_1)}-\frac{1}{|\tau_1-y_1|} \right)  \ln\frac{(\tau_1-y_1)^2}{y_1^2}\notag\\
	&~~~~+(\tau_1 \leftrightarrow \tau_2,\,y_1 \leftrightarrow y_2) \bigg\} _+ + 2\, (-3{\rm{L_z}}+2) \delta(\tau_1-y_1).	
\end{align}

\subsubsection{Matching kernel in momentum space}
The matching kernels for qGPDs in momentum space takes the following form in the $\overline{\rm MS}$ scheme:
\begin{itemize}
    \item Quark in quark
\end{itemize}
\begin{align}\label{eq:ms_exp_MOM}
	{\bb {C}}_{qq}^{\overline{\rm{MS}}}(x_1,x_2,y_1,y_2;{{\frac{\mu}{P_z}}})
	=&\delta(x_1-y_1)+a_sC_F \,{\bb {C}}_{qq}^{(1)}(x_1,x_2,y_1,y_2;{{\frac{\mu}{P_z}}})
\end{align}
with
\begin{align}\label{eq:quasikqq_MOM}
	{\bb {C}}_{qq,u}^{(1)}
	&= \bigg\{ \left( \frac{|x_1| }{y_1 (y_1+y_2)}(\rm{L_x}-1) +(x_1 \leftrightarrow x_2,y_1 \leftrightarrow y_2 )\right) -\frac{|x_1-y_1|}{y_1y_2}(\rm{L_{xy}}-1) \bigg\} +{\bb {C}}_{qq,t}^{(1)}\,  ,\notag\\
	{\bb {C}}_{qq,h}^{(1)}
	&={\bb {C}}_{qq,u}^{(1)}+ 2 \,\bigg\{  \frac{|x_1| }{y_1 (y_1+y_2)} +\frac{|x_2|}{y_2(y_1+y_2)} -\frac{|x_1-y_1|}{y_1y_2} \bigg\}  ,\notag\\	
	{\bb {C}}_{qq,t}^{(1)}
	&=\bigg\{ \left( \frac{|x_1| }{y_1 (y_1-x_1)}(\rm{L_x}-1) +(x_1 \leftrightarrow x_2,y_1 \leftrightarrow y_2 )\right)
	+\left(\frac{x_1}{y_1}+\frac{x_2}{y_2} \right)\frac{1}{|x_1-y_1|}(\rm{L_{xy}}-1)\bigg\} \, .
\end{align}

\begin{itemize}
    \item Quark in gluon
\end{itemize}
\begin{align}\label{eq:ms_exp_MOM2}
	{\bb {C}}_{qg}^{\overline{\rm{MS}}}(x_1,x_2,y_1,y_2;{{\frac{\mu}{P_z}}})
	=&-a_sT_FN_f \,{\bb {C}}_{qg}^{(1)}(x_1,x_2,y_1,y_2;{{\frac{\mu}{P_z}}})
\end{align}
with
\begin{align}\label{eq:quasi_Cqg}
	{\bb{C}}_{qg,u}^{(1)}&= \bigg\{ \left(\frac{|x_1| (2 x_2-y_2)}{ y_1^2 y_2 (y_1+y_2)}(\rm{L_x}-2) +\frac{2x_1 |x_1|}{y_1^2 y_2 (y_1+y_2)} -(x_1 \leftrightarrow x_2,y_1 \leftrightarrow y_2 ) \right)\notag\\
	&~~~+ \frac{(x_2-y_2)^2-x_1 x_2}{y_1^2y_2^2(x_1-y_1)} |x_1-y_1| \,\rm{L_{xy}} + \frac{2|x_1-y_1|(y_1-2x_1+x_2)}{y_1^2 y_2^2}\bigg\}\, ,\notag\\
	{\bb{C}}_{qg,h}^{(1)}&=\bigg\{ \left(\frac{|x_1|}{y_1^2 (y_1+y_2)}(\rm{L_x}-2) -(x_1 \leftrightarrow x_2,y_1 \leftrightarrow y_2 ) \right)+ \frac{2|x_1-y_1|(y_2-y_1)}{y_1^2 y_2^2} \notag\\
	&~~~+\frac{x_1(y_1-y_2)-y_1^2}{y_1^2y_2^2(x_1-y_1)}|x_1-y_1| \, \rm{L_{xy}}\bigg\}\, .
\end{align}

\begin{itemize}
    \item Gluon in quark
\end{itemize}
\begin{align}\label{eq:ms_exp_MOM3}
	{\bb {C}}_{gq}^{\overline{\rm{MS}}}(x_1,x_2,y_1,y_2;{{\frac{\mu}{P_z}}})
	=&-2a_sC_F \,{\bb {C}}_{gq}^{(1)}(x_1,x_2,y_1,y_2;{{\frac{\mu}{P_z}}})
\end{align}
with
\begin{align}\label{eq:quasi_Cgq_u}
	{\bb {C}}_{gq,u}^{(1)}= &\bigg\{\left( \frac{|x_1|(x_1-2y_1)}{y_1 (y_1+y_2)}\rm{L_x}+ \frac{2|x_1|(y_1-y_2)}{y_1(y_1+y_2)}-(x_1\leftrightarrow x_2,y_1\leftrightarrow y_2) \right) \notag\\
	&~~~+ \frac{|x_1-y_1|(x_1-3y_1+2y_2)}{y_1 y_2 } -\frac{(x_1-y_1)^2-y_1 y_2}{y_1 y_2(x_1-y_1)} |x_1-y_1| \,(\rm{L_{xy}}+1)\bigg\}\, ,\notag\\
	{\bb {C}}_{gq,h}^{(1)}= &\bigg\{ \left(\frac{-|x_1|x_1}{y_1 (y_1+y_2)} (\rm{L_x}-4) + \frac{|x_1|(y_1-2y_2)}{y_1(y_1+y_2)}-(x_1\leftrightarrow x_2,y_1\leftrightarrow y_2)\right)\notag\\
	&~~~+ \frac{2|x_1-y_1|(x_2-x_1)}{y_1y_2}  +\frac{(x_1-y_1)^2+y_1 y_2}{y_1 y_2(x_1-y_1)}|x_1-y_1| \,\rm{L_{xy}}\bigg\}\, .	
\end{align}

\begin{itemize}
    \item Gluon in gluon
\end{itemize}
\begin{align}\label{}
	{\bb {C}}_{gg}^{\overline{\rm{MS}}}(x_1,x_2,y_1,y_2;{{\frac{\mu}{P_z}}})
	=&\delta(x_1-y_1)+a_sC_A \,{\bb {C}}_{gg}^{(1)}(x_1,x_2,y_1,y_2;{{\frac{\mu}{P_z}}})
\end{align}
with
\begin{align}\label{eq:quasi_Kgg}
	{\bb{C}}_{gg,u}^{(1)}
	&=\bigg\{ \bigg(\frac{2 |x_1|x_1((y_1+y_2)(4y_1+y_2)-x_1(3y_1+y_2))}{ y_1^2 (y_1+y_2)^3}\left({\rm{L_x}}-\frac{5}{3} \right)+\frac{|x_1|x_1 (13 y_1+y_2)}{3 y_1^2 (y_1+y_2)^2}\notag\\
	&~~~~~~+(x_1\leftrightarrow x_2,\,y_1 \leftrightarrow y_2)\bigg)  -\frac{2|x_1-y_1|(x_1x_2+y_1y_2)}{y_1^2y_2^2} \left({\rm{L_{xy}}}-\frac{3}{2} \right)  \, \notag\\
       &~~~~~~-\frac{|x_1-y_1| \left((x_1-y_1)^2+12 y_1 y_2\right)}{3 y_1^2 y_2^2}\bigg\} +{\bb {T}}_{gg}^{(1)}, \notag\\
	{\bb{C}}_{gg,h}^{(1)}
	&=\, \bigg\{\left( \frac{ 2|x_1|x_1 }{y_1^2 (y_1+y_2)} \left({\rm{L_x}}-\frac{5}{2} \right)+ (x_1\leftrightarrow x_2,\,y_1 \leftrightarrow y_2)\right)-\frac{2|x_1-y_1|(x_1 y_2+x_2 y_1)}{y_1^2y_2^2}  \notag\\	
       &~~~~~~\times \left({\rm{L_{xy}}}-\frac{5}{2} \right)  -\frac{4|x_1-y_1|}{y_1 y_2} \bigg\}+{\bb {T}}_{gg}^{(1)},\notag\\
	{\bb{C}}_{gg,t}^{(1)}
	&=\, \bigg\{ \bigg( -\frac{|x_1|x_1 (x_1+3 x_2) (3y_1+y_2)}{3 {y_1}^2 (y_1+y_2)^3}  + \frac{4 |x_1|}{y_1(y_1+y_2)}+(x_1\leftrightarrow x_2,\,y_1 \leftrightarrow y_2)  \bigg) \notag\\
	&~~~~~~+\frac{|x_1-y_1|(-2x_1^2+y_1(y_1-9y_2)+x_1(y_1+3y_2))}{3y_1^2y_2^2} \bigg\} +{\bb {T}}_{gg}^{(1)} \, ,\notag\\
	{\bb {T}}_{gg}^{(1)}
	&= \, \bigg\{ \frac{x_1|x_1| }{y_1^2 (y_1-x_1)}\left({\rm{L_x}}-1 \right)
	+\frac{x_1^2}{y_1^2|x_1-y_1|}\left({\rm{L_{xy}}}-1 \right)+\frac{2(|x_1|-|x_1-y_1|)}{y_1^2}\notag\\
        &~~~~~~+(x_1\rightarrow x_2,y_1\rightarrow y_2) \bigg\}.
    \end{align}
In these formulas, we have defined
\beq
{\rm{L_x}}=\ln \frac{4{{P_z^2}} x_1^2}{\mu^2}\,, \qquad {\rm{L_{xy}}}=\ln \frac{4{{P_z^2}} (x_1-y_1)^2}{\mu^2}\,.
\eeq
We have checked that all evolution kernels in the above results are in agreement with Ref.~\cite{Belitsky:2005qn}. In addition, our results for $\bb {C}^{(1)}_{qq,\{h,t\}}, \bb {C}^{(1)}_{qg,\{h,t\}}$ are consistent with those in Ref.~\cite{Ma:2022ggj}, 
while the results for $\bb {C}^{(1)}_{qq,u}, \bb {C}^{(1)}_{qg,u}$ are slightly different from those of Ref.~\cite{Ma:2022ggj} due to different choices of quark operators. 
It is worth pointing out that one kinematic region is missing in the early calculation of the matching kernels in Refs.~\cite{Ji:2015qla,Xiong:2015nua,Liu:2019urm} due to an incomplete analytical continuation. To see this, we can temporarily assume $y>0$ in Eq.~(\ref{eq:quasikqq_MOM}), and do an explicit expansion of $\bb {C}_{qq}^{(1)}$ according to the regions of momentum fractions, 
then we have (by switching to the notation of Refs.~\cite{Ji:2015qla,Xiong:2015nua,Liu:2019urm} and assuming $\xi>0$)
\begin{align}\label{eq:quasikqqyexpanad1}
	\bb {C}_{qq}^{(1)}(x,y,\xi)
	=\frac{1}{y} &\bigg[ G_1(x,y,\xi) \theta(x<-\xi)\theta(x<y)+G_2(x,y,\xi) \theta(-\xi<x<\xi)\theta(x<y)\notag\\
	&+G_3(-x,-y,\xi){{\theta(-\xi<x<\xi)\theta(x>y)}}+G_3(x,y,\xi) \theta(x>\xi)\theta(x<y)\notag\\
	&-G_1(x,y,\xi) \theta(x>\xi)\theta(x>y)\bigg],
\end{align}
where the region $\theta(-\xi<x<\xi)\theta(x>y)$ is missing in the kinematic setup in Refs.~\cite{Ji:2015qla,Xiong:2015nua,Liu:2019urm}, and shall be recovered through a correct analytical continuation from the results there.

\subsection{Ratio and hybrid schemes}
So far our discussion has been focused on DR and $\overline{\rm MS}$ scheme. However, on the lattice where the lattice spacing acts as a UV regulator, a different renormalization scheme has to be adopted. A convenient choice in the literature is the ratio renormalization~\cite{Radyushkin:2018cvn}, which is based on the fact that the quark and gluon quasi-LF operators are multiplicatively renormalized~\cite{Ji:2015jwa,Chen:2016fxx,Ishikawa:2017faj,Ji:2017oey,Green:2017xeu,Zhang:2018diq,Li:2018tpe,Braun:2020ymy}, thus all UV divergences in the bare qLFCs can be removed by dividing by the correlation of the same operator in a zero-momentum hadronic state. The ratio scheme does not depend on the regularization, therefore, the matching in this scheme can be conveniently obtained from the results in DR and $\overline{\rm MS}$ scheme by dividing by the corresponding zero momentum matrix element result. The ratio scheme matching can be directly applied to lattice renormalized qLFCs (after taking the continuum limit of the latter) at short distances. 

However, the ratio renormalization cannot be directly used in the LaMET factorization approach. In this approach, one needs to do a Fourier transform to momentum space, which also requires the qLFCs at large distances. The ratio renormalization factor (i.e., the zero momentum matrix element) at such distances introduces undesired IR contributions that can invalidate the factorization formula. To avoid this, one turns to a hybrid renormalization scheme~\cite{Ji:2020brr} where one renormalizes the bare qLFCs at short distances using ratio renormalization, while renormalizes the bare qLFCs at large distances via the Wilson line mass renormalization~\cite{Chen:2016fxx} or self renormalization~\cite{LatticePartonCollaborationLPC:2021xdx}. In this way, the renormalization at large distances removes only the UV divergences and associated renormalon terms, but does not introduce additional IR effects.

In this section, we present the matching results in the ratio and hybrid schemes.

\subsubsection{Matching in coordinate space}
The ratio renormalization is done by forming the following ratio of qLFCs
\begin{align}\label{eq:nonsinglet_ratiofun}
{\cal{H}}(P_z,\mu^2 z_{12}^2)/{\cal{H}}(0,\mu^2 z_{12}^2),
\end{align}
where the denominator is the qLFC at zero momentum. To obtain the ratio scheme matching, we need to calculate the denominator in a parton state also to the NLO. 
The results read
\begin{itemize}
    \item Quark in quark
\end{itemize}
\begin{align}\label{eq:zeromomqq}
	{\cal{H}}_{qq}(0,\mu^2 z_{12}^2)&=Z_{q}(\mu^2 z_{12}^2)=Z_{q}^{(0)}+Z_{q}^{(1)}(\mu^2 z_{12}^2)=1+2a_sC_F \left(-A_{4}\,\rm{L_z}+A_{5} \right)\, ,\\
	A_{4,u}&=\frac{3}{2}\,, \qquad A_{4,h}=\frac{3}{2}\, , \qquad A_{4,t}=2\, ,\notag\\
	A_{5,u}&=\frac{5}{2}\,, \qquad A_{5,h}=\frac{7}{2}\, , \qquad A_{5,t}=2\, .\notag
\end{align}
\begin{itemize}
    \item Gluon in gluon
\end{itemize}
\begin{align}\label{eq:zeromomgg}
	{\cal{H}}_{gg}(0,\mu^2 z_{12}^2)&=Z_{g}(\mu^2 z_{12}^2)=Z_{g}^{(0)}+Z_{g}^{(1)}(\mu^2 z_{12}^2)=1+2a_sC_A \left(-E_{3}\,\rm{L_z}+E_{4} \right)\, ,\, \\
	E_{3,u}&=\frac{11}{6}\,, \qquad E_{3,h}=\frac{7}{3}\, , \qquad E_{3,t}=3\, ,\notag\\
	E_{4,u}&=4\,, \qquad E_{4,h}=\frac{7}{3}\, , \qquad E_{4,t}=\frac{7}{2}\, .\notag
\end{align}
At one-loop, the ratio scheme matching is related to the $\overline{\rm MS}$ scheme matching as follows:
\begin{align}\label{eq:ratiofun_exp}
	 C^{\text{ratio}}(\alpha,\beta,\mu^2 z_{12}^2)
	 =&\,\,C^{\overline{\rm{MS}}}(\alpha,\beta,\mu^2 z_{12}^2)-Z^{(1)}(\mu^2 z_{12}^2)\delta(\alpha)\delta(\beta).
\end{align}
For example, 
for $C_{qq}^{\text{ratio}}$ we have
\begin{align}
	C_{qq}^{\text{ratio}}&(\alpha,\beta,\mu^2 z_{12}^2)\notag\\
	&=
	2a_sC_F \bigg\{ \left(A_{2} + \frac{\bar{\alpha}}{\alpha}\,\delta(\beta) + \frac{\bar{\beta}}{\beta} \,\delta(\alpha) \right)(\rm{L_z}-1)+A_{3}-2\frac{\ln\alpha}{\alpha}\delta({\beta})-2\frac{\ln\beta}{\beta}\delta({\alpha})\bigg\}_+ 
\end{align}
with the coefficients $A_i$ given in Eq.~(\ref{eq:coo_C_qq}).
Note that the one-loop matching kernel {{(diagonal element)}} in the ratio scheme is a complete {{plus-}}distribution as expected from the definition of the scheme.

Now we turn to the hybrid scheme, where the renormalized qLFCs take the following form~\cite{Ji:2020brr}
\begin{align}\label{eq:hybrid}
	C^{\text{hybrid}}&\left(\alpha,\beta,\mu^2 z_{12}^2,z_{12},z_s\right)\notag\\
	&=C^{\text{ratio}}(\alpha,\beta,\mu^2 z_{12}^2)\,\theta\left(z_s-|z_{12}|\right) +C^{\overline{\rm{MS}}} (\alpha,\beta,\mu^2 z_{12}^2) \,{{e^{-\delta m |z_{12}|}}} \, {\rm{Z}}_h(z_s)\,\theta\left(|z_{12}|-z_s\right),
\end{align}   
where the Wilson line mass renormalization factor ${{e^{-\delta m |z_{12}|}}}$ contains a linearly divergent counterterm $\delta m$ to remove the linear UV divergence arising from the Wilson line self-energy in lattice regularization, which is absent in DR. $z_s$ denotes a truncation point within the perturbative region and ${\rm{Z}}_h(z_s)$ ensures that the renormalized qLFCs are continuous at ${{z_{12}=z_s}}$. This gives,
\begin{align}
	{\rm{Z}}_h(z_s)=\frac{e^{\delta m |z_s|}}{Z(\mu^2z_s^2)}.
\end{align}
The one-loop matching kernel in the hybrid scheme can be obtained from that in the ratio scheme by adding an extra term
 \begin{align}	C^{\text{hybrid}}\bigg(\alpha,\beta,\mu^2 z_{12}^2,\frac{z_{12}^2}{z_s^2}\bigg)=C^{\text{ratio}}(\alpha,\beta,\mu^2 z_{12}^2) -T_L \ln \frac{z_{12}^2}{z_s^2}\delta(\alpha) \delta(\beta) \,\theta\left(|z_{12}|-z_s\right),
 \end{align}
 where $T_L$ is the coefficient of $\rm{L_z}$ in the zero-momentum qLFCs in Eq.~(\ref{eq:zeromomqq}) for ${C}_{qq}$ and $C_{qg}$ and in Eq. (\ref{eq:zeromomgg}) for ${C}_{gq}$ and $C_{gg}$. 
 
 \subsubsection{Matching in momentum space}
After a Fourier transform with respect to the quasi-LF distance, we obtain the following ratio scheme matching in pseudo space, which is related to that in the $\overline{\rm MS}$ scheme 
\begin{align}\label{eq:ratiofun_exp_pseudo}
	{\cal C}^{\text{ratio}}(\tau_1,\tau_2 , y_1,y_2;{\mu^2 z_{12}^2})
	=&{\cal C}^{\overline{\rm{MS}}}(\tau_1,\tau_2 , y_1,y_2;{\mu^2 z_{12}^2})-Z^{(1)}\left(\mu^2 z_{12}^2\right)\, \delta(\tau_1-y_1).
\end{align}
From Eq.~\eqref{eq:ratiofun_exp_pseudo}, it is straightforward to obtain the matching for qGPDs in momentum space by the Fourier transform in Eq.~\eqref{eq:quasiGPDdef}. The result reads
\begin{align}\label{eq:ratiofun_exp_pseudo_mom}
	{\bb C}^{\text{ratio}}\left(x_1,x_2 , y_1,y_2;\frac{\mu}{ P_z}\right)
	=&{\bb C}^{\overline{\rm{MS}}}\left(x_1,x_2 , y_1,y_2;\frac{\mu}{ P_z}\right)-T_L\,\frac{1}{|x_1-y_1|}.
\end{align}
The hybrid scheme matching can be obtained from this by adding an extra term resulting from the truncation in Eq.~(\ref{eq:hybrid})
 \begin{align}\label{eq:hybrid_Cqq_mom}
 	{\bb C}^{\text{hybrid}}&\left(x_1,x_2 , y_1,y_2;\frac{\mu}{ P_z}\right)\notag\\
  &={\bb C}^{\text{ratio}}\left(x_1,x_2 , y_1,y_2;\frac{\mu}{ P_z}\right)-T_L\left[ -\frac{1}{|x_1-y_1|}+\frac{2\rm{Si}((x_1-y_1)\lambda_s)}{\pi(x_1-y_1)}  \right]_+ .
 \end{align}

\section{PDF and DA limits}
\label{SEC:pdfdalimits}
In this section, we present the matching kernels for the PDFs and DAs that emerge from the results in previous sections by taking special kinematic limits.

\subsection{PDFs}
In coordinate space, the factorization formula for the forward qLFCs defining the PDFs takes the form
\begin{align}\label{}
\tilde h(z_{12},{{p_z}},\mu)= 
\int_0^1 d\alpha\,\bm{C}(\alpha,\mu^2 z_{12}^2) h^{l.t.}({{\bar{\alpha}}},\mu),
\end{align}
with the following matching kernels
\begin{itemize}
    \item Quark in quark
\end{itemize}
\begin{align}
\bm{C}_{qq}^{(1)}(\alpha,\mu^2z_{12}^2)=&2a_sC_F \bigg\{ \left( \alpha \,A_{2} + 2\left[\frac{\bar{\alpha}}{\alpha}\right]_+ \right)\left(\rm{L_z}-1\right)+\alpha \, A_{3} -4\left[\frac{\ln(\alpha)}{\alpha}\right]_+   \bigg\}\,\notag\\
&+2a_sC_F \left(-2\,\rm{L_z}+2\right)\delta(\alpha)\,.
\end{align}
\begin{itemize}
    \item Quark in gluon
\end{itemize}
\begin{align}
 \bm{C}^{(1)}_{qg}(\alpha,\mu^2z_{12}^2)&=4ia_sT_F N_f \,{\mathbf z}_{12} \,B_5\, {\rm{L_z}} \, ,\notag\\
B_{5,u}&=\frac{2}{3}\alpha^3-\alpha^2+\alpha\,, \qquad
B_{5,h}=\alpha(1-\alpha).
\end{align}
\begin{itemize}
    \item Gluon in quark
\end{itemize}
\begin{align}
\bm{C}_{gq}^{(1)}(\alpha,\mu^2z_{12}^2)& = \frac{-2ia_sC_F}{{\mathbf z_{12}}}\bigg\{ \alpha\,D_3 \left(\rm{L_z}+1\right)-4\bar{\alpha} +\left(\rm{L_z}+D_4\right)\delta(\alpha) \bigg\}\, .
\end{align}
\begin{itemize}
    \item Gluon in gluon
\end{itemize}
{\begin{align}
 \bm{C}^{(1)}_{gg}(\alpha,\mu^2z_{12}^2)=&2a_sC_A \bigg\{ \left(E_{5} + 2\left[\frac{\bar{\alpha}^2}{\alpha}\right]_+ \right)\left({\rm{L_z}}-1\right)+E_{6} -4\left[\frac{\ln(\alpha)}{\alpha}\right]_+  \bigg\}\,\notag\\
  &+2a_sC_A \left(-3\,{\rm{L_z}}+2\right)\delta(\alpha)\,, \notag
  \end{align}
  \begin{align}
&E_{5,u}=2\alpha\,(2-2\alpha+\alpha^2)\, , \qquad  E_{5,h}=4\alpha(1-\alpha)\, , \qquad  E_{5,t}=0 ,\notag\\
&E_{6,u}=2 \alpha\, (5-5\alpha+3\alpha^2 )\, , \qquad  E_{6,h}=\frac{3}{2} \,E_{5,h}\, , \qquad  E_{6,t}=-\frac{2}{3}\,\alpha (\alpha^2-3\alpha-3) .
\end{align}}
The coefficients $A_{2,3}$ and $D_{3,4}$ are given in Eq.~(\ref{eq:coo_C_qq}) and Eq.~(\ref{eq:Cgq-pos}). Here the matching coefficients $\bm{C}_{qq}^{(1)}$ are consistent with those in Refs.~\cite{Izubuchi:2018srq, Braun:2021gvv}. 
 
 In the literature, the short-distance factorization is often carried out in coordinate space instead of the pseudo space. Therefore, we skip the pseudo-PDFs here and consider the momentum space factorization for the quasi-PDFs (qPDFs) only. 
The complete factorization for quark and gluon PDFs takes the form\footnote{Note that we follow the conventions in Ref.~\cite{Belitsky:2005qn} throughout. Other definitions for the LF-GPDs/PDFs may result in factorization formulas of slightly different forms with $x g(x)\Big|_{\text{our work}} \mapsto g(x)\Big|_{\text{others}}$. }
\begin{align}\label{eq:PDF_FF}
	&\tilde q\left(x,\frac{\mu}{P_z}\right)=\int_{-1}^{1} dy\, \left[{F}_{qq}\left(x,y;{{{\frac{\mu}{P_z}}}}\right)q(y,\mu)+ {F}_{qg}\left(x,y;{{{\frac{\mu}{P_z}}}}\right)yg(y,\mu)\right],\notag\\
 & x\tilde g\left(x,\frac{\mu}{P_z}\right)=\int_{-1}^{1} dy\, \left[ {F}_{gq}\left(x,y;{{{\frac{\mu}{P_z}}}}\right)q(y,\mu)+ {F}_{gg}\left(x,y;{{{\frac{\mu}{P_z}}}}\right)yg(y,\mu)\right].
\end{align}
 Since we have assumed $t=0$ throughout our calculation, the forward limit can be simply obtained by taking the skewness $\xi \rightarrow 0$. 
The matching kernels of qPDFs are obtained from that of qGPDs through
\beq
{F}\left(x,y;{{{\frac{\mu}{P_z}}}}\right)=\bb {C}\left(x,-x,\, y,-y;{{{\frac{\mu}{P_z}}}}\right).
\eeq
It is clear that no reduction formula for gluon transversity case in a spin-one-half hardon. For hardons with spin one or higher, gluon transversity is visible in the forward limit~\cite{Belitsky:2005qn,Diehl:2003ny}
The results for the singlet PDFs are summarized below.
\begin{itemize}
    \item Quark in quark
\end{itemize}
\begin{align}\label{eq:quasikqq_forward1}
	{F}_{qq,u}^{(1)}
	&=2a_sC_F  \, \bigg\{ \frac{(y-x)}{2y^2} \frac{|x|}{x} \left(\rm{l_{x}}-1\right)+\frac{|y-x|}{2y^2} \left(\rm{l_{xy}}-1\right) +\frac{1}{y}\frac{|x|}{x}  \bigg\} +	{F}_{qq,t}^{(1)}\,  ,\notag\\
	{F}_{qq,h}^{(1)}
	&={F}_{qq,u}^{(1)}+ 4a_s C_F \,\bigg\{  \frac{(y-x)}{2y^2} \frac{|x|}{x} +\frac{|y-x|}{2y^2}   \bigg\}  ,\notag\\
	 {F}_{qq,t}^{(1)}
	&=2a_sC_F \, \bigg\{\frac{|x|}{y(y-x)}\left(\rm{l_{x}}-1\right)+\frac{x}{y|x-y|} \left(\rm{l_{xy}}-1\right) \bigg\} \,,
\end{align}
\begin{itemize}
    \item Quark in gluon
\end{itemize}
\begin{align}\label{eq:quasikqq_forward2}
	{F}_{qg,u}^{(1)}
	&= -2a_sT_F N_f \, \bigg\{\frac{((x-y)^2+x^2)}{2y^4} \,\left(-\frac{|x|}{x}\,\rm{l_{x}}+\frac{|x-y|}{x-y}\,\rm{l_{xy}}\right) + \frac{3x(|x|-|x-y|)}{y^4} \notag\\
      &~~~~~~~~~~~~~~~~~~+ \frac{-2y|x|+y|x-y|}{y^4}\bigg\},\notag\\
	{F}_{qg,h}^{(1)}
	&=-2a_sT_F N_f \, \bigg\{\frac{(2x-y)}{2y^3} \left(-\frac{|x|}{x}\,\rm{l_{x}}+\frac{|x-y|}{x-y}\,\rm{l_{xy}}\right)+ \frac{2(|x|-|x-y|)}{y^3}\bigg\},
\end{align}
\begin{itemize}
    \item Gluon in quark
\end{itemize}
\begin{align}\label{eq:quasikqq_forward3}
	{F}_{gq,u}^{(1)}
	&= - 4a_sC_F \, \bigg\{\frac{((x-y)^2+y^2)}{2y^2} \,\left(-\frac{|x|}{x}\,\rm{l_{x}}+\frac{|x-y|}{x-y}\,\rm{l_{xy}}\right) + \frac{|x-y|}{2(x-y)}- \frac{|x|}{y}+ \frac{2|x-y|}{y} \bigg\}, \notag\\
	{F}_{gq,h}^{(1)}
	&=-4a_sC_F  \, \bigg\{ \frac{(x-2y)}{2y^2} \,\left(|x|\,\rm{l_{x}}-\frac{x|x-y|}{x-y}\,\rm{l_{xy}}\right) + \frac{4x^2|x-y|+|x|(3y^2+2xy-4x^2)}{2 xy^2} \bigg\},
\end{align}
\begin{itemize}
    \item Gluon in gluon
\end{itemize}
\begin{align}\label{}
F_{gg,u}^{(1)}
&=2a_s C_A\bigg\{-\frac{ |x|(x-y)(x^2+y^2)}{x y^4}\, {\rm{l_x}} +\frac{ |x-y|(x^2+y^2)}{y^4} \,{\rm{l_{xy}}}+\frac{|x|(5x^3-3xy^2+6y^3)}{3x y^4}\notag\\
&~~~~+\frac{|x-y|(-5x^2+xy+y^2)}{3y^4}\bigg\} +F_{T}^{(1)},\notag\\
F_{gg,h}^{(1)}
&=2a_s C_A\bigg\{ \frac{2|x|(y-x)}{y^3}\, \left({\rm{l_x}}-\frac{5}{2}\right) +\frac{2x|x-y|}{y^3}\, \left({\rm{l_{xy}}}-\frac{5}{2}\right)+\frac{2(|x|+|x-y|)}{y^2}\bigg\} +F_{T}^{(1)},\notag\\
F_{gg,t}^{(1)}
&=2a_s C_A\bigg\{\frac{x^2+x y-5y^2}{3xy^4}\left(|x|(x-y)-x|x-y|\right)\bigg\}+F_{T}^{(1)} ,\notag\\
F_{T}^{(1)}
&=2a_s C_A\bigg\{\frac{x|x|}{(y-x)y^2}\, \left({\rm{l_x}}-1\right) +\frac{x^2}{|x-y|y^2}\, \left({\rm{l_{xy}}}-1\right)+\frac{2\left(|x|-|x-y|\right)}{y^2}\bigg\}\,,
\end{align}
where 
\beq
{\rm{l_x}}=\ln \frac{4{{P_z^2}} x^2}{\mu^2}\,, \qquad {\rm{l_{xy}}}=\ln \frac{4{{P_z^2}} (x-y)^2}{\mu^2}\,.
\eeq
Several remarks are in order. First of all, our results for $F_{qq}^{(1)}$ are consistent with those in Ref.~\cite{Chou:2022drv}. For the mixing matching coefficients $F_{qg}^{(1)}$,
our result for the polarized case is consistent with that in Ref.~\cite{Ma:2022ggj}, while for the unpolarized case there is a slight difference because we have chosen a different Dirac structure in the quark bilinear operator. Secondly, our ratio scheme result for the unpolarized gluon PDF is in agreement with that derived in Eq. (7.28) of Ref.~\cite{Balitsky:2019krf}}, which also starts from coordinate space and then Fourier transforms to momentum space. However, both the results of Ref.~\cite{Balitsky:2019krf} and our results show a discrepancy with earlier calculations performed directly in momentum space~\cite{Wang:2019tgg} in the unphysical region. {In the ratio scheme, the contribution of the unphysical region in momentum space comes entirely from the Fourier transform of the $\ln z_{12}^2\mu^2$ term in coordinate space. From this perspective, it is obvious that the momentum space matching in the unphysical region is entirely determined by the evolution kernel in coordinate space, where the latter is well-known in the literature. Nevertheless, it is highly desirable to further investigate and understand what leads to the mismatch between these two approaches.

\subsection{DAs}
Finally, we consider the kinematic limit where the DAs are recovered. The quark operators with $\Gamma=\gamma^z \gamma_5, \gamma^t, \gamma^t\gamma^\perp\gamma_5$ that we introduce at the beginning define the leading Fock state quasi-DA in a pseudoscalar, longitudinally polarized vector, and transversely polarized vector meson, respectively. We denote them by subscripts $p$, $h$ and $t$. In coordinate space, the factorization formula for the DAs takes the same form as that for the GPDs, and the matching kernels are the same as shown in Eq.~(\ref{eq:coo_C_qq}). In momentum space, the factorization formula becomes
\begin{align}\label{}
\tilde{\phi}\left(x,\frac{\mu}{P_z}\right)=\int_{0}^{1} dy\, V\left(x,y;{{{\frac{\mu}{P_z}}}}\right)\phi(y,\mu),
\end{align}
where the matching kernel $V(x,y)$ can be obtained from that for the qGPDs by taking the limit
\begin{align}\label{limit_DA}
V\left(x,y;{{{\frac{\mu}{P_z}}}} \right)=\bb {C}\left(x,1-x,y,1-y;{{{\frac{\mu}{P_z}}}}\right),
\end{align}
where the quark and antiquark momentum fractions are given by $x$ $(y)$ and $1-x$ $(1-y)$, respectively. For completeness, we summarize the results of the matching kernel below,
\begin{align}\label{eq:quasikqq_forward4}
	V_{qq,h}^{(1)}
	&=a_sC_F  \, \bigg\{ \frac{|x|}{y}  \left({\rm{l_x}}-1\right)  +\frac{|1-x|}{(1-y)}\left({\rm{l_{\bar{x}}}}-1\right) + \frac{|x-y|}{y(y-1)} \left({\rm{l_{xy}}}-1\right)  \bigg\} +	V_{qq,t}^{(1)}\,  ,\notag\\
	V_{qq,p}^{(1)}
	&=V_{qq,h}^{(1)}+ 2a_s C_F \,\bigg\{  \frac{|x|}{y}    +\frac{|1-x|}{1-y}  + \frac{|x-y|}{(y-1)y}  \bigg\}  ,\notag\\	
	V_{qq,t}^{(1)}
	&=a_sC_F \,  \bigg\{ \frac{|x|}{y(y-x)}  \left({\rm{l_x}}-1\right) +\frac{|1-x|}{(1-y)(x-y)}  \left({\rm{l_{\bar{x}}}}-1\right)+ \frac{x+y-2xy}{|x-y|y(1-y)} \left({\rm{l_{xy}}}-1\right)  \bigg\} \,,
\end{align}
where we have defined
\beq
{\rm{l_{\bar{x}}}}=\ln \frac{4{{P_z^2}} (1-x)^2}{\mu^2}.
\eeq
If we expand the above expressions according to the regions of momentum fractions, we obtain
\begin{align}\label{eq:quasikqqyexpanad2}
	V_{qq}^{(1)}(x,y)
	= &H_1(x,y) \theta(x<0<y)+H_2(x,y) \theta(0<x<y)
	+H_2(1-x,1-y) \theta(y<x<1)\notag\\
	&+H_1(1-x,1-y) \theta(y<1<x),
\end{align}
which are in agreement with existing results in the literature~\cite{Ji:2015qla,Liu:2018tox,Xu:2018mpf}. 

\section{Conclusion}
\label{SEC:conclusion}
In this paper, we have developed a unified framework for perturbative calculations of the hard-matching kernel connecting collinear qLFCs to the LF correlations. We started by deriving the matching kernel for flavor-singlet quark and gluon correlations in non-forward kinematics in DR and $\overline{\rm MS}$ scheme in coordinate space. The results for the GPDs, PDFs and DAs were obtained by choosing appropriate kinematics. We then converted our results to the state-of-the-art ratio and hybrid schemes, and gave the matching kernel both in coordinate and in momentum space. Our results provide a complete manual for a state-of-the-art extraction of
all leading-twist GPDs, PDFs as well as DAs from lattice simulations, both for the flavor-singlet and nonsinglet combinations, and either in coordinate or in momentum space factorization approach. Our framework has the potential to greatly facilitate higher-order perturbative calculations involving collinear qLFCs. We will push this forward and present the NNLO results in forthcoming publications.

\vspace{2em}
{\noindent\bf Note added:} While this paper is being finalized, another {{preprint~\cite{Ma:2022gty}}} appears which calculates the one-loop matching for gluon quasi-GPDs in momentum space with a completely different approach. We find that the results agree up to terms that vanish upon convolution
with the lightcone distributions (and thus have no impact on physical results), except for
the mixing term $C_{gq}$ which we believe is due to the prescription used in dealing with the
additional pole $1/z_{12}$ that compensates for the mismatch in the mass dimensions of the
gluon and quark fields. While our prescription ensures correct Mellin moments to all orders,
the prescription of~\cite{Ma:2022gty} does not. We will investigate this issue in a future publication.


\acknowledgments
We thank Jian-Ping Ma and Zhuoyi Pang for helpful discussions.
This work is supported in part by National Natural Science Foundation of China under grants No. 11975051, No. 12061131006 (FY and JHZ), and by the Collaborative Research
Center TRR110/2 funded by the Deutsche Forschungsgemeinschaft (DFG, German Research Foundation) under grant 409651613 (Y.J.). 

\appendix

\section{Technical details about the Fourier transform}
\label{sec:app1}
As mentioned in the main text, there exist certain ambiguities in the Fourier transform of the mixing terms. In this Appendix, we give a more detailed discussion on this.
\subsection{Gluon in quark}
\label{app:Cgq}
We begin with the mixing channel ${\cal C}_{gq}$. In the first place, we have the qLFCs in coordinate space,
\begin{align}\label{eq:coo_qlfcs}
{\cal{H}}_{gq}^{(1)} 
&= \frac{-2ia_sC_F\Gamma(-\epsilon)\eta^\epsilon}{{\mathbf z_{12}}}\,\bigg\{ D_1\, O_q(z_1,z_2) +2 \epsilon \, D_1 \int^1_0d\alpha \left[O_q(z_1,z_{21}^\alpha)+O_q(z_{12}^\alpha,z_2)\right]\notag\\
&~~~~~~~~~~~~~~~~~~~~~~~~~~~+ 2\,D_2\int^1_0d\alpha\int^{\bar\alpha}_0 d\beta\, O_q(z_{12}^\alpha,z_{21}^\beta)\bigg\}\, .
\end{align}
The definition of the pGPDs and the factorization formula have been given in Eqs.~(\ref{eq:pseudoFT}) and~(\ref{eq:pseudoFactor}) in the main text. It is relatively simple to relate the qLFCs and the pGPDs as $z_{12}^2$ is fixed in the Fourier transform, which reads
\begin{align}
{\cal{C}}_{gq,1}^{(1)}=-\int^{\tau_1}_{\infty(-1+i\epsilon)} d \tau_1 \bigg\{&2a_sC_F \Gamma(-\epsilon)\eta^\epsilon \, \bigg[D_1\delta(\tau_1-y_1)+2D_2 \int^1_0 d\alpha d\beta \, \bar{\alpha}\, \delta(\tau_1-\bar{\alpha}y_1-\bar{\alpha}\beta y_2)\notag\\
&+ 2\epsilon\,D_1\int^1_0 d\alpha \Big( \delta(\tau_1-\bar{\alpha}y_1)+ \delta(\tau_2-\bar{\alpha}y_2) \Big) \bigg]\bigg\},
\end{align} 
where ${-i\int^{\tau_1}_{\infty(-1+i\epsilon)} e^{-i\tau_1 z_{12}} d \tau_1 =\frac{1}{z_{12}} e^{-i\tau_1 z_{12}}}$, and the momentum-fraction conservation $\tau_1+\tau_2=y_1+y_2$ is implied. Two formulas in Ref.~\cite{Ji:2014eta} are helpful in deriving the coefficient functions for pGPDs from their quasi-LF counterparts,
\begin{align}\label{}
\int_0^1 d\alpha f(\alpha) \delta(x-\alpha y)=f\left(\frac{x}{y}\right)\, {\vartheta}_{11}^0(x,x-y)
\end{align}
and
\begin{align}
\int_0^1 d\alpha \bar{\alpha}^n {\vartheta}_{11}^0(x_1-y_1 \bar\alpha,x_1-\eta \bar\alpha)=\frac{1}{n} &\bigg\{ \left[ 1-\left(\frac{x_1}{\eta}\right)^n\right] \bb{\vartheta}_{11}^0(x_1-y_1,-x_2)\notag\\
&-\frac{y_1}{y_2} \left[\left(\frac{x_1}{\eta}\right)^n-\left(\frac{x_1}{y_1}\right)^n\right] {\vartheta}_{11}^0(x_1-y_1,x_1)\bigg\}\, ,
\end{align}
where ${\vartheta}_{11}^0(x_1,x_2)$ is defined in terms of the usual step function, specifically,
\beq
{\vartheta}_{11}^0(x_1,x_2)=\frac{\theta(x_1)-\theta(x_2)}{x_1-x_2}.
\eeq
One can smoothly go to the pseudo-space following these two formulas and making full use of the momentum-fraction conservation. The pseudo matching coefficient gives
\begin{align}
{\cal{C}}_{gq,2}^{(1)}
=&-2a_sC_F \Gamma(-\epsilon)\eta^\epsilon \bigg\{ D_2 \left[\frac{\tau_1^2 \, \theta(\tau_1)}{y_1(y_1+y_2)}+\frac{\tau_2^2 \, \theta(-\tau_2)}{y_2(y_1+y_2)}-\frac{(\tau_1-y_1)^2}{y_1 y_2} \theta(\tau_1-y_1)\right] \notag\\
&+D_1\theta(\tau_1-y_1)+ 2\epsilon\,D_1 \left[\frac{\tau_1}{y_1}\,\theta(\tau_1)+ \frac{\tau_2}{y_2}\,\theta(-\tau_2) +\frac{(\tau_1-y_1)(y_1-y_2)}{y_1 y_2}\theta(\tau_1-y_1) \right]\bigg\}.
\end{align} 
The ambiguities in the lower limit of the $ \tau_1$ integral are removed  by matching the Mellin moments computed in coordinate and pseudo space. In coordinate space, the (bare) coefficient function~\eqref{eq:coo_qlfcs} generates the Mellin moments as
\begin{align}
{\cal{H}}_{gq}^{(1)} [O_q(z_1,z_2)]=\gamma_{gq}^{(1)} O_g(z_1,z_2)
\end{align} 
with $O_q(z_1,z_2)=z_{12}^j$ and $O_g(z_1,z_2)=i z_{12}^{j-1}$, we have
\begin{align}
\gamma_{gq}^{(1)}=-2a_sC_F \Gamma(-\epsilon)\eta^\epsilon  \left\{ D_1+\frac{2 D_2}{(1+j)(2+j)}+\frac{4\,\epsilon \,D_1}{1+j} \right\}.
\end{align} 
Since it is clear that only the $\theta(x_1)$ and its symmetric channel $\theta(-x_2)$ harbor ambiguities ($\theta(x_1-x_2)$ channel actually has no contribution), we can take the
following ansatz for the matching coefficient in the pseudo space,
\begin{align}\label{eq:kernelmatch}
\mathcal C^{(1)}_{gq,3}
=&-2a_sC_F \Gamma(-\epsilon)\eta^\epsilon \bigg\{ 2\epsilon \, D_1\left[\frac{(b_1 \tau_1 y_1+b_2 \tau_1 y_2)\,\theta(\tau_1) }{y_1(y_1+y_2)}+ \frac{(b_1 \tau_2 y_2+b_2 \tau_2 y_1)\,\theta(-\tau_2) }{y_2(y_1+y_2)} \right]  \notag\\
&+D_2 \left[\frac{(a_1 \tau_1^2 +a_2 \tau_1 y_1+a_3 y_1^2)  \, \theta(\tau_1)}{y_1(y_1+y_2)}+\frac{(a_1 \tau_2^2 +a_2 \tau_2 y_2+a_3 y_2^2)  \, \theta(-\tau_2)}{y_2(y_1+y_2)}\right]\bigg\},
\end{align} 
where we have used  the symmetry property of ${\mathcal C}^{(i)}$ under $\tau_1 \leftrightarrow \tau_2$, $y_1 \leftrightarrow y_2$.
The anstaz for ${\cal{C}}_{gq}^{(1)}$ follows from a simple scaling property, which gives another ansatz expressed as two variable functions of $\tau/\xi$ and $y/\xi$,
\begin{align}
 {\cal C}_{gq}^{(i)}\left(\xi+\tau,\xi-\tau,\xi+y,\xi-y \right)= {\scr C}_{gq}^{(i)}\left(\frac{\tau}{\xi},\,\frac{y}{\xi}\right).
\end{align} 
Now, we apply the following identity,
\begin{align}
\underset{\xi\rightarrow \,0}\lim\, \int_0^1 dx x^{j-1}\,\scr C_{gq}^{(i)}\left(2\,\frac{x}{\xi}-1,\,\frac{2}{\xi}-1\right)=-\frac{\gamma_{gq}^{(i)}(j)}{j},\,\qquad j\geq 1.
\end{align}
In addition, the relation $\scr C_{gq}^{(i)}(\omega_1,\omega_2)=-\scr C_{gq}^{(i)}(-\omega_1,-\omega_2)$ applies. 
Finally, we can solve the algebraic equation w.r.t. the parameters ($a_i,b_i$) such that~Eq.~\eqref{eq:kernelmatch} is satisfied for arbitrary $j$,  hence finally yielding the expression in the pseudo space,
\begin{align}\label{eq:kernelmatch2}
{\cal C}_{gq,4}^{(1)}
=&-2a_sC_F \Gamma(-\epsilon)\eta^\epsilon \bigg\{ D_1 \, \theta(\tau_1-y_1)+D_2 \bigg[\frac{(\tau_1^2 +a_2 \tau_1 y_1) \, \theta(\tau_1)}{y_1(y_1+y_2)}+\frac{(\tau_2^2 +a_2 \tau_2 y_2)  \, \theta(-\tau_2)}{y_2(y_1+y_2)}\notag\\
& -\frac{(\tau_1-y_1)^2}{y_1 y_2} \theta(\tau_1-y_1) \bigg] 
+ 2\epsilon \, D_1 \bigg[\frac{(b_1 \tau_1 y_1+ \tau_1 y_2)\,\theta(\tau_1) }{y_1(y_1+y_2)}+ \frac{(b_1 \tau_2 y_2+ \tau_2 y_1)\,\theta(-\tau_2) }{y_2(y_1+y_2)} \notag\\
&+\frac{(\tau_1-y_1)(y_1-y_2)}{y_1 y_2}\theta(\tau_1-y_1)\bigg]\bigg\}
\end{align} 
with 
\begin{align}
a_{2,u}=-2\,, \qquad a_{2,h}&=0\, , \qquad b_{1,u}=-2\,, \qquad b_{1,h}=-\frac12\, . \qquad \notag
\end{align} 

\subsection{Quark in gluon}

The mixing channel ${\cal C}_{qg}$ includes some structures that have no influence to physical results. Starting form Eq.~\eqref{eq:Ccalqg} in coordinate space, it is straightforward to give the pseudo space coefficients,
\begin{align}
{\cal{C}}_{qg,1}^{(1)}
&\!=\!-4a_sT_FN_f\Gamma(-\epsilon) {\eta^\epsilon}\delta(\tau_1+\tau_2-y_1-y_2)\partial_{\tau_1} \int^1_0d\alpha\int^{\bar\alpha}_0 d\beta (\bar{\alpha}\bar{\beta}+3\alpha\beta) \delta(\tau_1-\bar{\alpha}y_1-\beta y_2)\notag\\
&\!=\!-4a_sT_F\Gamma(-\epsilon) {\eta^\epsilon}\, \bigg\{ \frac{\tau_1[2\tau_2(3y_1+y_2)-(y_1+y_2)^2]}{y_1^2(y_1+y_2)^3} \theta(\tau_1)\notag\\
&~~~~~~~+\frac{2\tau_1^2+y_1^2-\tau_1(3y_1+y_2)}{y_1^2y_2^2} \theta(\tau_1-y_1)+
\frac{\tau_2[2\tau_1(3y_2+y_1)-(y_1+y_2)^2]}{y_2^2(y_1+y_2)^3} \theta(-\tau_2)\bigg\}\, .
\label{}
\end{align}
On the other hand, it is easy to check that the matching kernel Eq.~\eqref{eq:pGPD_qg} properly reproduces the Mellin moments of the position space following the procedure spelled out in Appendix~\ref{app:Cgq}, 
\begin{align}
{\cal{C}}_{qg,2}^{(1)}
=-4a_sT_FN_f\Gamma(-\epsilon) {\eta^\epsilon}\, &\bigg\{ \frac{\tau_1(2\tau_2-y_2)}{y_1^2y_2(y_1+y_2)} \theta(\tau_1)+\frac{2\tau_1^2+y_1^2-\tau_1(3y_1+y_2)}{y_1^2y_2^2} \theta(\tau_1-y_1)\notag\\
&+
\frac{\tau_2(2\tau_1-y_1)}{y_1y_2^2(y_1+y_2)} \theta(-\tau_2)\bigg\}\, .
\label{eq:pGPD_qg}
\end{align}
meaning that it is indeed also the correct moment space matching coefficient. It implies that in general, we are free to add certain terms to $\theta(\tau_1)$ and $\theta(-\tau_2)$ without affecting the Mellin moments. 
Such additional terms are, therefore ``unphysical'' in the sense that they do not generate any changes to physical
quantities, such as IR-subtraction or GPD evolution. One particular choice reads,
\begin{align}
\delta \mathcal{C}_{qg}^{(1)}=-4a_s T_F N_f \Gamma(-\epsilon)\left\{ \frac{2\tau_1\tau_2(y_1-y_2)}{y_1y_2(y_1+y_2)^3}[\theta(\tau_1)-\theta(-\tau_2)]\right\}\, .
\end{align} 

\section{Unpolarized gluon operator with d-dimension}
We also consider the situation that the indices $\mu,\nu$ of the $O_{g,u}$ operator can be taken over all Lorentz components~(d-dimension). The matching Kernels for $O_{g,u}$ are summarized below.

\subsection{In coordinate space}
\begin{itemize}
    \item Gluon in quark
\end{itemize}
\begin{align}\label{}
  C_{gq}^{\overline{\rm{MS}}}(\alpha,\beta,\mu^2 z_{12}^2) &= \frac{-2ia_sC_F}{{\mathbf z_{12}}}\bigg\{ \Big(\delta(\alpha)\delta(\beta)+2\Big) \left({\rm{L_z}}+1\right)+6-2\Big(\delta(\alpha)+\delta(\beta)\Big) \bigg\}\, ,\notag\\
\end{align}
\begin{itemize}
    \item Gluon in gluon
\end{itemize}
\begin{align}
 C_{gg}^{\overline{\rm{MS}}}(\alpha,\beta,\mu^2 z_{12}^2)& = 
 \delta(\alpha)\delta(\beta)
 +2a_sC_A \bigg\{ \left(e_1 + \left[\frac{\bar{\alpha}^2}{\alpha}\right]_+\delta(\beta) + \left[\frac{\bar{\beta}^2}{\beta}\right]_+ \delta(\alpha) \right)\left({\rm{L_z}}-1\right)+e_2\notag\\ &-2\left[\frac{\ln(\alpha)}{\alpha} \right]_+\delta({\beta})-2\left[\frac{\ln(\beta)}{\beta}\right]_+\delta({\alpha})   \bigg\}+2a_sC_A \left(-3\,{\rm{L_z}}+2\right)\delta(\alpha)\delta(\beta)\,, \notag\\
e_1&=4(1-\alpha-\beta+3\alpha\beta)\,, \qquad e_2=3\,e_1+8\alpha\beta\, . \qquad 
\end{align} 
 
\subsection{In pseudo space}
\begin{itemize}
    \item Gluon in quark
\end{itemize}
\begin{align}\label{}
	{\cal{C}}_{gq}^{\overline{\rm{MS}}}(\tau_1,\tau_2,y_1,y_2;\mu^2 z_{12}^2)
	=&-2a_sC_F\,{\cal{C}}^{(1)}_{gq}(\tau_1,\tau_2,y_1,y_2;\mu^2 z_{12}^2)
\end{align}
with
\begin{align}\label{}
	{\cal{C}}_{gq,u}^{(1)}
	=&\,\bigg\{  \left(\frac{|\tau_1|(\tau_1-2 y_1)}{y_1(y_1+y_2)} -\frac{(\tau_1-y_1)^2-y_1 y_2}{y_1 y_2(\tau_1-y_1)} |\tau_1-y_1|-\frac{|\tau_2|(\tau_2-2 y_2)}{y_2(y_1+y_2)}\right) \left({\rm{L_z}+1}\right) \notag\\
	&~~~+\frac{|\tau_1|(3\tau_1-y_1-2y_2)}{y_1(y_1+y_2)} -\frac{ |\tau_1-y_1|(3\tau_1-y_1-2y_2)}{y_1 y_2}-\frac{|\tau_2|(3\tau_2-2y_1-y_2)}{y_2(y_1+y_2)}\bigg\}\,,\notag\\
\end{align}
\begin{itemize}
    \item Gluon in gluon
\end{itemize}
\begin{align}\label{}
	{\cal{C}}_{gg}^{\overline{\rm{MS}}}(\tau_1,\tau_2,y_1,y_2;\mu^2 z_{12}^2)
	=&\delta(\tau_1-y_1)+a_sC_A\,{\cal{C}}^{(1)}_{gg}(\tau_1,\tau_2,y_1,y_2;\mu^2 z_{12}^2)
\end{align}
with
\begin{align}\label{}
	{\cal{C}}_{gg,u}^{(1)}
	&= \bigg\{ \frac{ 2|\tau_1|\tau_1((y_1+y_2)(4y_1+y_2)-\tau_1(3y_1+y_2))}{ y_1^2 (y_1+y_2)^3}\left({\rm{L_z}}+\frac83\right)-\frac{4|\tau_1|\tau_1}{3y_1^2(y_1+y_2)}\notag\\
	&~~~~-\frac{|\tau_1-y_1|(\tau_1\tau_2+y_1y_2)}{y_1^2y_2^2} \left({\rm{L_z}}+2\right)+\frac{2|(\tau_1-y_1)^3|}{3y_1^2y_2^2}+(\tau_1 \leftrightarrow \tau_2,\,y_1 \leftrightarrow y_2)\bigg\}+{\cal{T}}_{gg}^{(1)} \,  ,\notag\\
	{\cal{T}}_{gg}^{(1)}
	&=\bigg\{ \left(\frac{\tau_1|\tau_1|}{y_1^2(y_1-\tau_1)}+\frac{\tau_1^2}{y_1^2|\tau_1-y_1|} \right)({\rm{L_z}}-1)  
	+\left(\frac{|\tau_1|}{\tau_1(\tau_1-y_1)}-\frac{1}{|\tau_1-y_1|} \right)  \ln\frac{(\tau_1-y_1)^2}{y_1^2}\notag\\
	&~~~~+(\tau_1 \leftrightarrow \tau_2,\,y_1 \leftrightarrow y_2) \bigg\} _+ + 2\, (-3{\rm{L_z}}+2) \delta(\tau_1-y_1).	
\end{align}
\subsection{In momentum space}
\begin{itemize}
    \item Gluon in quark
\end{itemize}
\begin{align}\label{}
	{\bb {C}}_{gq}^{\overline{\rm{MS}}}(x_1,x_2,y_1,y_2;{\mu}/{P_z})
	=&-2a_sC_F \,{\bb {C}}_{gq}^{(1)}(x_1,x_2,y_1,y_2;{\mu}/{P_z})
\end{align}
with
\begin{align}\label{}
	{\bb {C}}_{gq,u}^{(1)}= &\bigg\{\left( \frac{|x_1|(x_1-2y_1)}{y_1 (y_1+y_2)}\rm{L_x}+ \frac{|x_1|(x_1+y_1-2y_2)}{y_1(y_1+y_2)}-(x_1\leftrightarrow x_2,y_1\leftrightarrow y_2) \right) \notag\\
	&~~~ + \frac{2|x_1-y_1|(y_2-y_1)}{y_1 y_2}-\frac{(x_1-y_1)^2-y_1 y_2}{y_1 y_2(x_1-y_1)} |x_1-y_1|  \,(\rm{L_{xy}}+1)\bigg\}\, .\notag\\
\end{align}

\begin{itemize}
    \item Gluon in gluon
\end{itemize}
\begin{align}\label{}
	{\bb {C}}_{gg}^{\overline{\rm{MS}}}(x_1,x_2,y_1,y_2;\mu/P_z)
	=&\delta(x_1-y_1)+a_sC_A \,{\bb {C}}_{gg}^{(1)}(x_1,x_2,y_1,y_2;{\mu}/{P_z})
\end{align}
with
\begin{align}\label{}
	{\bb{C}}_{gg,u}^{(1)}
	&=\bigg\{ \bigg(\frac{2 |x_1|x_1((y_1+y_2)(4y_1+y_2)-x_1(3y_1+y_2))}{ y_1^2 (y_1+y_2)^3}\left({\rm{L_x}}-1 \right)+\frac{4|x_1|x_1 }{ y_1 (y_1+y_2)^2}\notag\\
	&~~~~~~+(x_1\leftrightarrow x_2,\,y_1 \leftrightarrow y_2)\bigg)  -\frac{2|x_1-y_1|(x_1x_2+y_1y_2)}{y_1^2y_2^2} \left({\rm{L_{xy}}}-1 \right)  -\frac{4|x_1-y_1|}{y_1 y_2}\bigg\} +{\bb {T}}_{gg}^{(1)}, \notag\\
	{\bb {T}}_{gg}^{(1)}
	&= \, \bigg\{ \frac{x_1|x_1| }{y_1^2 (y_1-x_1)}\left({\rm{L_x}}-1 \right)
	+\frac{x_1^2}{y_1^2|x_1-y_1|}\left({\rm{L_{xy}}}-1 \right)+\frac{2(|x_1|-|x_1-y_1|)}{y_1^2}\notag\\
        &~~~~~~+(x_1\rightarrow x_2,y_1\rightarrow y_2) \bigg\}.
    \end{align}


\vspace{2em}


\bibliographystyle{apsrev}
\bibliography{ref}

\end{document}